\documentclass[aps,reprint,showpacs,superscriptaddress,
footinbib,citeautoscript]{revtex4-1}

\usepackage{graphicx}

\usepackage[utf8]{inputenc}
\usepackage{csquotes}
\usepackage[american]{babel}
\usepackage[T1]{fontenc}
\usepackage{enumerate}
\usepackage{mdwlist}
\usepackage[activate=normal]{pdfcprot}
\usepackage{bbding}
\usepackage{color}
\usepackage{amssymb}
\usepackage{amsmath}
\usepackage{amsfonts}
\usepackage{mathrsfs}
\usepackage{lipsum}
\usepackage{upgreek}

\usepackage{bm}
\usepackage{dcolumn}
\usepackage{color}
\usepackage[colorlinks=true,citecolor=blue]{hyperref}
\hypersetup{colorlinks=true,citecolor=blue,linkcolor=red
,urlcolor=blue}

\begin{document}
\title{Plasmon-phonon-polaritons in encapsulated phosphorene}


\author{Farnood G. Ghamsari}
\email{ghamsari@ipm.ir}
\affiliation{Department of Physics, Kharazmi University, Tehran, 15719-14911, Iran}
\affiliation{School of Physics, Institute for Research in Fundamental Sciences (IPM), Tehran 19395-5531, Iran}
\author{Reza Asgari}
\affiliation{School of Physics, Institute for Research in Fundamental Sciences (IPM), Tehran 19395-5531, Iran}
\affiliation{School of Nano Science, Institute for Research in Fundamental Sciences (IPM), Tehran 19395-5531, Iran}
%
\date{\today}
%
\begin{abstract}
We consider a system consists of a doped monolayer phosphorene embedded between two hexagonal Boron Nitride (hBN) slabs along the heterostructure direction. The wavevector azimuthal angle dependence of the plasmon-polariton and plasmon-phonon-polariton modes of the hybrid system are calculated based on the random phase approximation at finite temperature. The collective modes illustrate strong anisotropy and strong coupling with phonon modes of the polar media and furthermore, the Landau damping occurs due to the intraband processes when plasmon enters intraband electron-hole continuum. Our numerical results show that the plasmon mode is highly confined to the surface along the zigzag direction. Owing to the strong electron-phonon interaction, the phonon dispersions in the Reststrahlen bands are also angle-dependent. These results are also in agreement with those of the semiclassical model obtained in our calculations.
\\
\linebreak
\end{abstract}
\maketitle
\section{Introduction}
\label{intro}
\begin{figure}
\hspace*{-0.22cm}
   \includegraphics[width=0.614\linewidth]{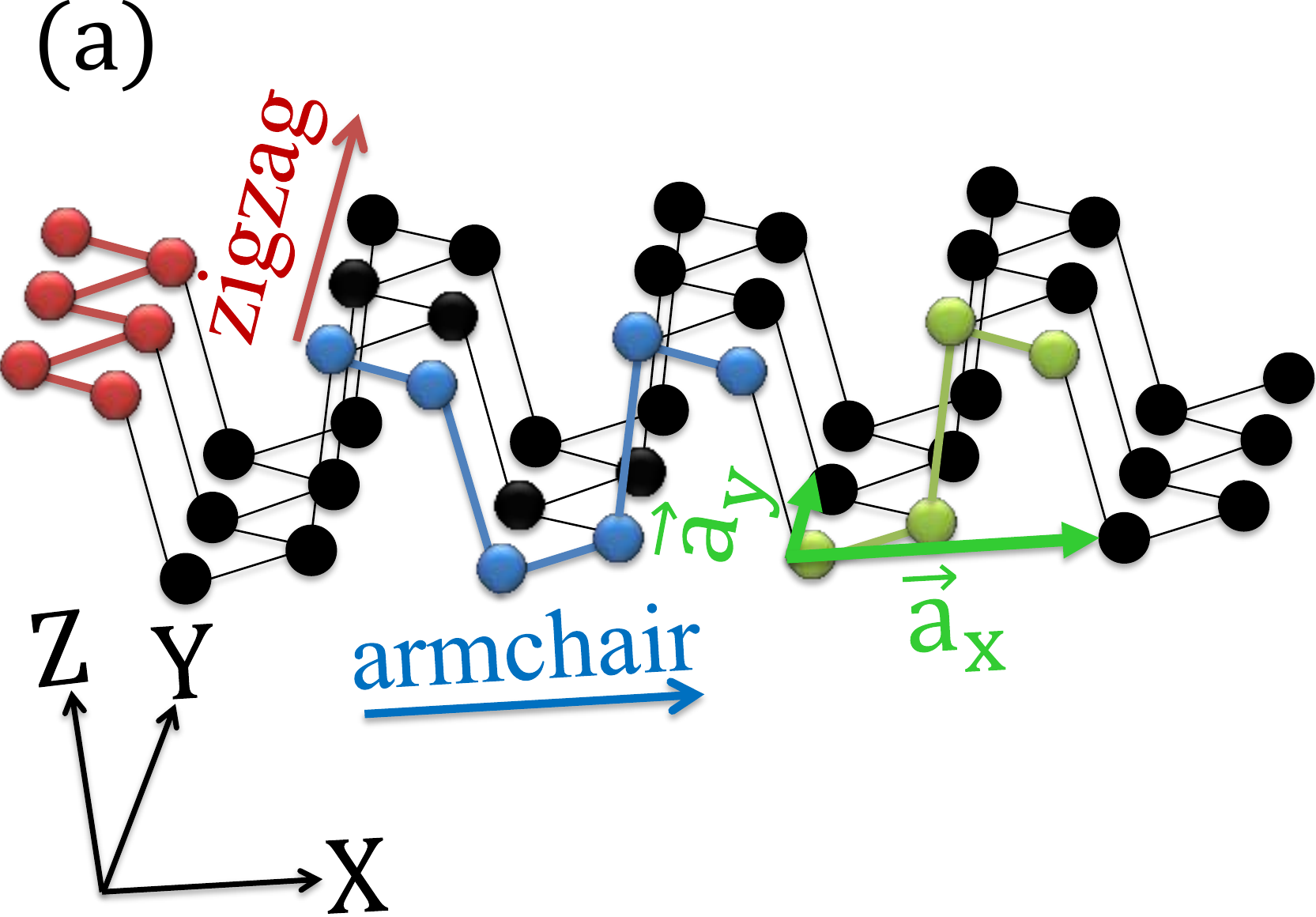}
   \includegraphics[width=0.385\linewidth]{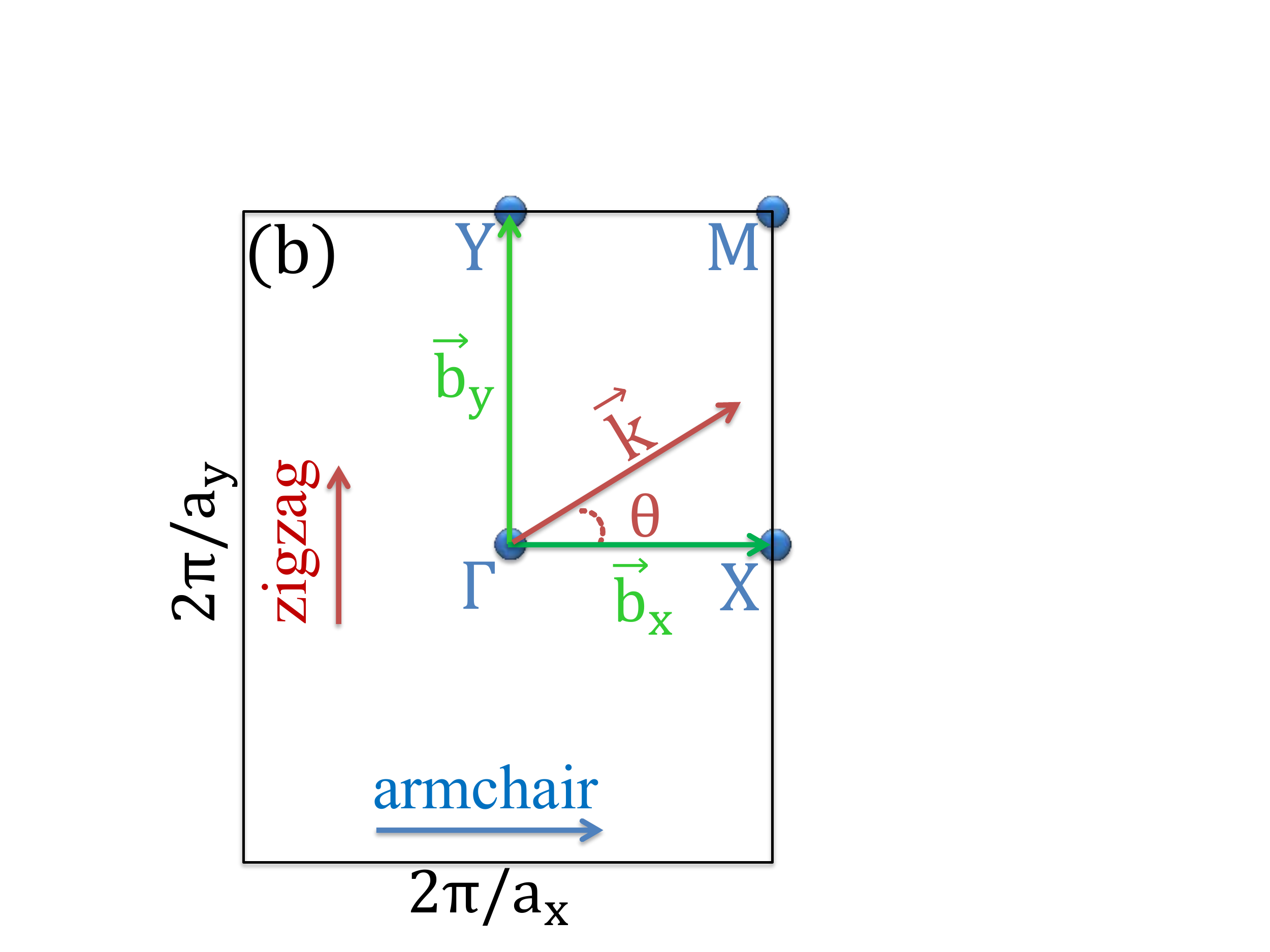}\\
   \hspace*{-0.28cm}
   \includegraphics[width=0.492\linewidth]{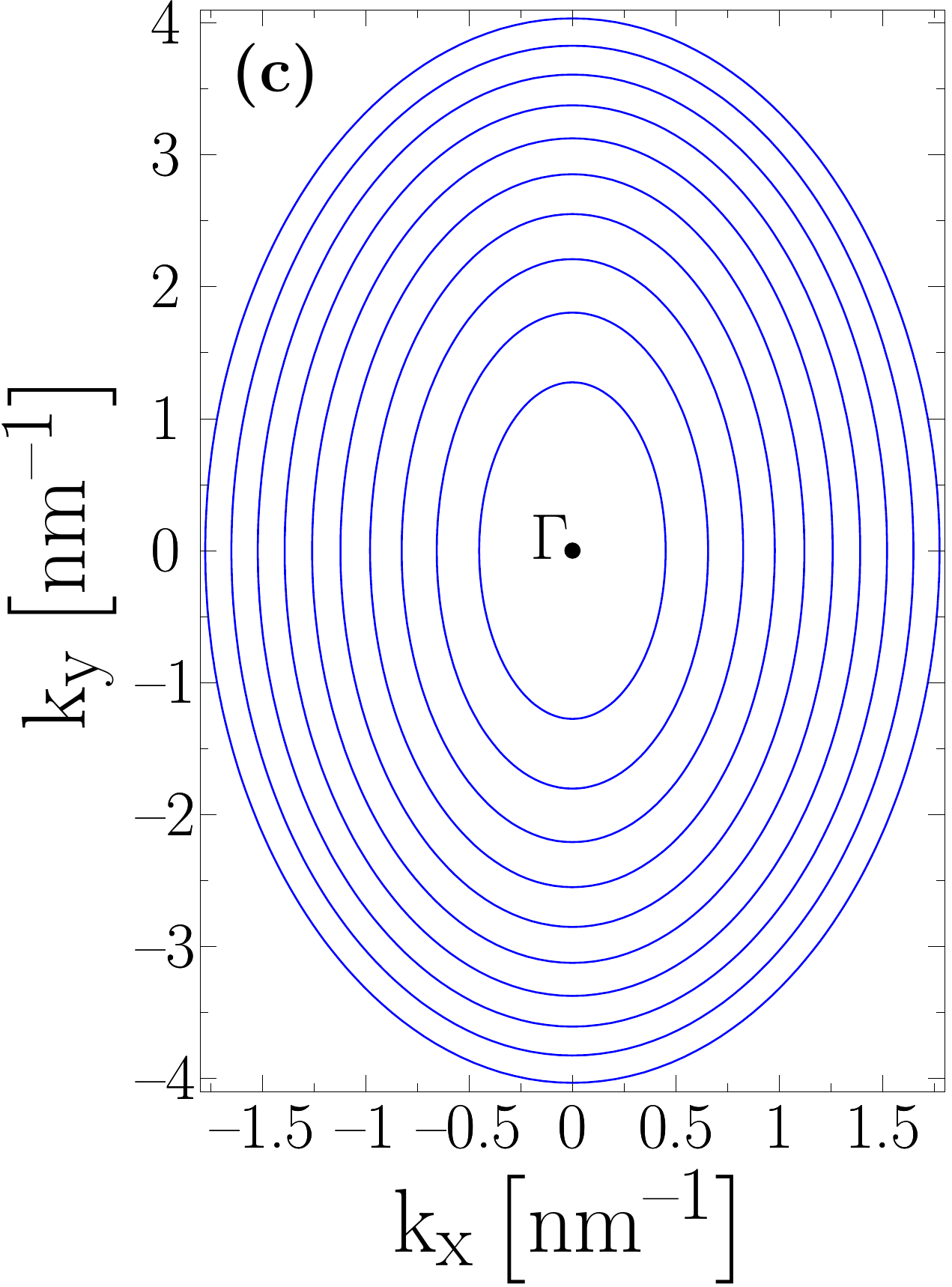}
   \includegraphics[width=0.499\linewidth]{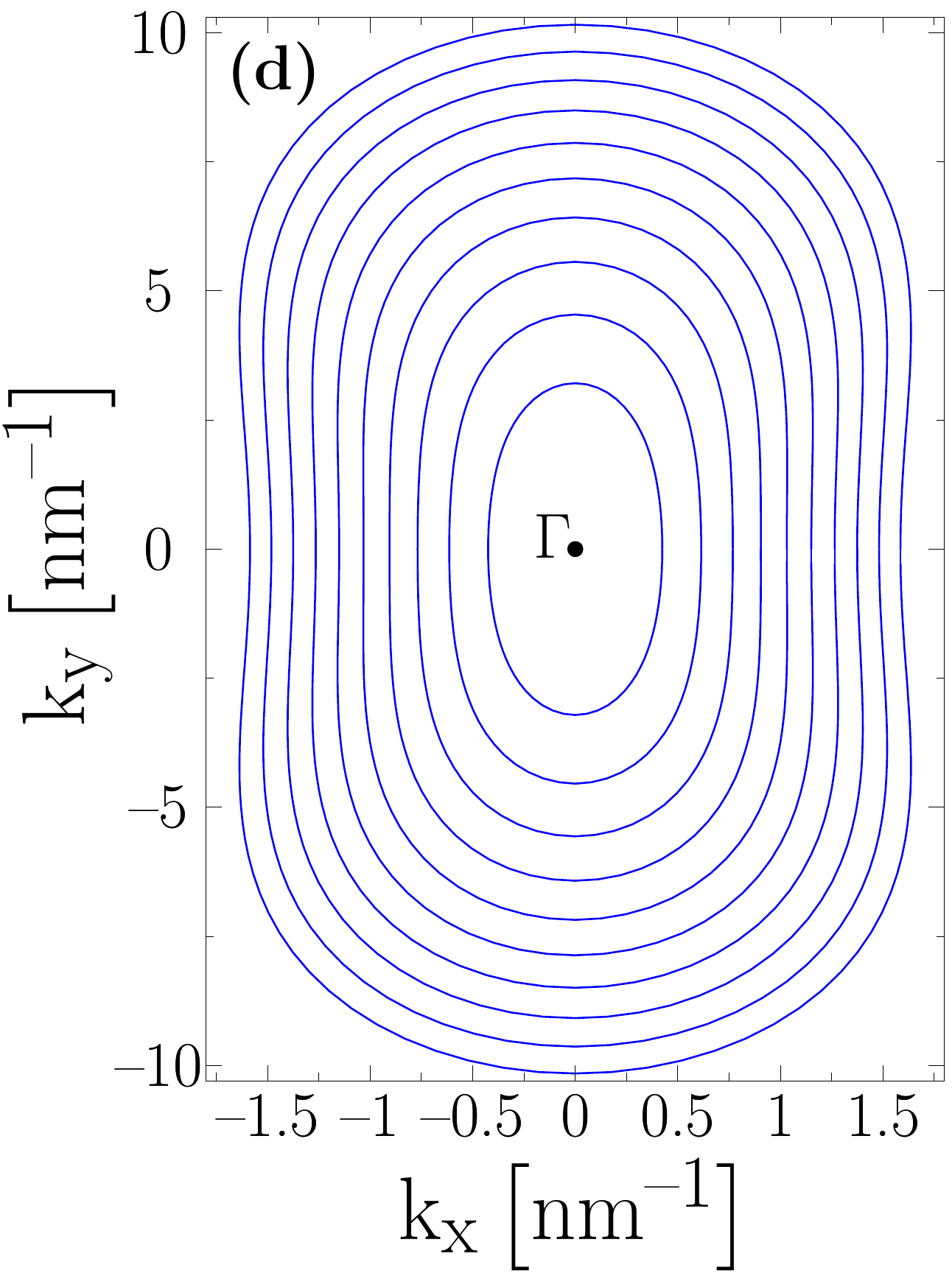}
   \caption{(Color online) (a) A schematic picture of phosphorene structure. The black colored atoms show the whole orthorhombic structure of the phosphorene crystal. The green colored atoms form the unit cell of phosphorene defined by the primitive vectors $\mathrm{{\bf a}_x}$, $\mathrm{{\bf a}_y}$. The blue and red colored atoms depict the armchair and the zigzag edges respectively. (b) The first Brillouin zone of phosphorene defined by the reciprocal lattice primitive vectors $\mathrm{{\bf b}_x}$, $\mathrm{{\bf b}_y}$ in addition to labels of high symmetry points. Wavevector ${\bf k}$ and  its azimuthal angle $\theta$ are also illustrated. Isofrequency contour surface of (c) electrons and (d) holes in the $\mathrm{k}$-space, for different values of the Fermi energy $\mathrm{0<\varepsilon_F<0.5}$ eV with step 0.05 eV at $\mathrm{T=0 \, K}$ are shown. Both (c) and (d) have a same scale along $\mathrm{k_x}$-axis, but they have different scales along $\mathrm{k_y}$-axis; the valence band (d) is about 3 times more dispersed in zigzag direction in comparison with the conduction band (c).}
   \label{fig1}
\end{figure}

%
Phosphorene is the monolayer counterpart of Black Phosphorus (BP); a single-atomic layered material consisting of only phosphorus atoms \cite{Bridgman,Keyes} with five outer shell electrons. In phosphorene, the phosphorus atoms are tightly packed in a rectangular lattice with the structure being slightly puckered - see Fig. \ref{fig1}(a) - giving rise to novel correlated electronic properties ranging from semiconducting to superconducting behaviors \cite{Asahina,Sugai,Liu14,Li,Tran,Xia,Qiao}. But unlike graphene (as a well-known 2D material), this puckered structure of phosphorene impose a strong anisotropy in the band structure and therefore in the collective excitations like the standard plasmon mode and its extensions such as surface plasmon-phonon-polariton modes in the related heterostructures described in the following.

In order to trigger and measure plasmon modes, a scanning near-field optical microscope can be used in which the aperture radius is much smaller than the wavelength of incident light. The required in-plane momenta is provided by the near-field evanescent components of light coming out from the microscope. For example, plasmon modes in doped graphene \cite{Garcia,Jablan,Goncalves,Tame} have been measured by using this technique with nanometer resolution. Two research groups carried out experiments on graphene plasmonics using a similar technique \cite{Fei,Chen}.
\begin{figure}[t]
   \centering
   \hspace*{-0.2cm}
   \includegraphics[width=\linewidth]{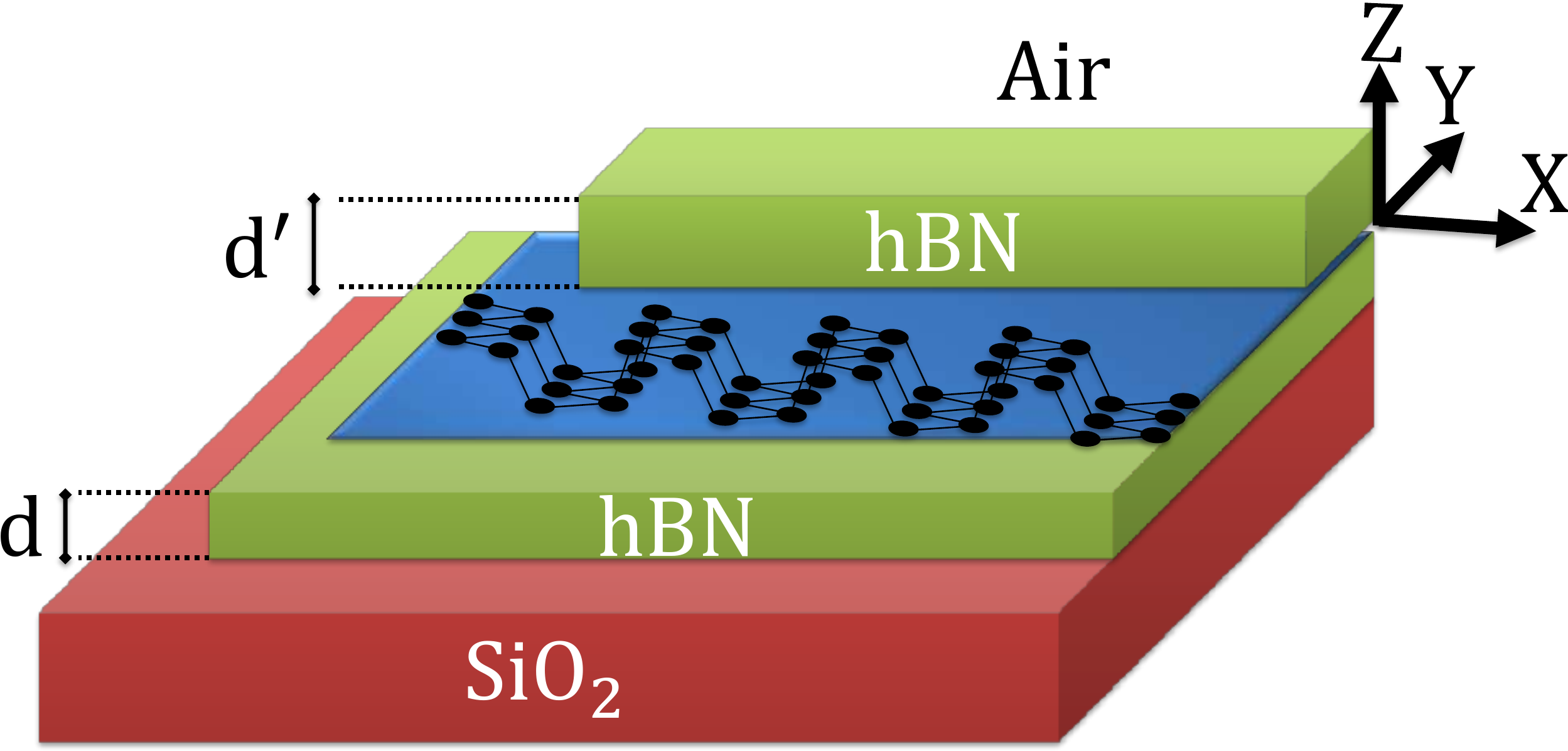}
   \caption{(Color online) A schematic picture of phosphorene encapsulated between two hBN slabs (with thicknesses $\mathrm{d}$ and $\mathrm{d'}$) along the heterostructure direction (z-axis) on a $\mathrm{SiO_2}$ semi-infinite substrate (media 1). The encapsulated phosphorene (media 2) is in contact with the air from above (media 3).} 
 	\label{fig2}
\end{figure}

Phonon-polaritons, on the other hand, are collective modes originating from the coupling of photons with optical phonons in polar dielectric materials \cite{Poddubny,Caldwell}. It is well-known that in the longitudinal optical phonon mode, the two different charged ions move in anti-phase and create a dipole and therefore, the electric field of the dipole can interact with electrons through the long-range Coulomb interaction which is known as polar coupling \cite{Mahan}. Plasmon-phonon-polarition arises when a two-dimensional (2D) crystalline material is deposited on a polar substrate such as SiO$_2$ or hexagonal Boron Nitride (hBN). Using Raman scattering \cite{Mooradian} in electron doped GaAs, the first plasmon-phonon coupling measurement was reported and the plasmon-phonon coupling in graphene has been explored and addressed by many groups \cite{Liu,Kock,Luxmoore}.

Surprisingly, hBN with a weak Van der Waals bonded nature, which is a natural hyperbolic material, supports tunable propagating phonon-polaritons in the bulk \cite{Caldwell,Dai15,Tomadin}. The dielectric permittivities of such materials assume unique forms at the phonon-polariton modes. In between frequencies, the hBN material behaves like a conventional metal with a negative dielectric constant and strongly reflect incident radiation. The range of frequencies for which light propagation is forbidden is called the Reststrahlen band. This forbidden band has been observed for a unique phonon-polariton medium of hBN  \cite{Dai14,Brar}. Combining hBN with 2D crystalline structure gives rise to unconventional plasmon-phonon hybridization and this hybrid system can be used for tailoring novel subwavelength metamaterials.

As there are various possibilities for the dispersion of the electromagnetic wave penetrating and propagating in hBN ranging from normal elliptic, gapped/gapless hyperbolic up to indefinite ones, it is highly potent to reveal extraordinary behaviors within the realms of plasmonics, nanophotonics and subwavelength engineering \cite{Zhou}. New studies uncover the fact that hBN's phonon-polariton are not exclusively belong to its bulk, since some modes propagating on its edges with highly confined field have been detected \cite{Zhao}. Moreover, some recent theoretical studies have shown that bending hBN thin films would marvelously emerge out new edge phonon-polariton modes propagating along particular directions with relatively low losses that raises it as a new alternative to nanowires and nanoribbons within the phonon-polariton-based nanophotonic engineering route \cite{Zhou}. hBN's interesting optical phenomena are not limited to these progresses; it has tunable plasmon-phonon interaction \cite{Brar,Dai15,Woessner,Xu,Kumar} in addition to the fact that its ability to work as a subwavelength wave guide with controllable focusing is studied \cite{Dai14,XuGhamsari,Caldwell,Dai15b} and its potential for very high resolution imaging is shown \cite{Li15}.

Strong coupling between hBN Fabry-P\'{e}rot phonon-polariton modes and the collective charge oscillations in a doped 2D crystalline materials' sheet gives rise to hybrid excitation surface plasmon-phonon-polaritons (SPPP) which reshapes their standard plasmon modes. As an example, SPPP modes in graphene on hBN has been recently measured \cite{Dai15,Woessner} by using infrared transmission techniques. The hybridization creates two new plasmon-phonon-polariton modes with dispersion relations that are distinctly different from the original graphene plasmon dispersion. As mentioned above, these modes are isotropic, but this article shows that their corresponding modes in phosphorene are in high influence of its anisotropic band structure and are no longer isotropic; instead, they have a strong angle-dependence. 

In this paper, we focus on a system consists of a monolayer phosphorene embedded between two hBN slabs along the heterostructure direction. We use a low-energy model Hamiltonian together with the Random-Phase Approximation (RPA) to investigate the charge excitations spectrum of doped phosphorene in the presence of different dielectric media.  Having calculated the density-density response function, we can therefore calculate the macroscopic dielectric function whose imaginary part gives the optical absorption spectrum, and the collective modes are established by the zero in the real part of the macroscopic dielectric function (more precisely, by its pole). Here, we are just interested in the low-energy excitations for investigating the collective modes.

We first illustrate an isofrequency contour surface in the $k$-space to explore their symmetries in both the electron and hole doped cases. Having calculated the Lindhard function at finite temperature, the azimuthal angle dependence of the plasmon-polariton and plasmon-phonon-polariton modes of the hybrid system are calculated. Two significant features visible in the results are (1) the complete reshaping of the typical square root dispersion of the standard plasmon modes to a three-branched dispersion of SPPP modes in the vicinity of the frequency of the hBN's optical phonon, and more importantly, (2) the highly anisotropic nature and angle dependence of these new emerged-out trifurcated modes. The collective modes, which are decomposed by the electron-phonon interaction, illustrate strong anisotropic and strong coupling with phonon modes of the polar media, and, moreover, the Landau damping occurs due to the intraband processes when plasmon enters intraband electron-hole continuum. The lowest plasmon dispersion is displayed as a $\sqrt q$ originating from the long-wavelength limit of charge carrier's scattering. Owing to the strong electron-phonon interaction, the phonon dispersions in the Reststrahlen bands are angle-dependent too. Furthermore, we show that the energy of the collective modes, which also depends on the Fermi energy, decreases by increasing the wavevector azimuthal angle and the SPPP modes are phonon-like in the zigzag direction. We also obtain the semiclassical model of the collective behaviors for an anisotropic system in the long-wavelength limit based on the classical electrodynamics and show that the RPA results are in agreement with those of the latter.

The paper is organized as follows. In Sec.~\ref{sec2}, we present the model Hamiltonian, calculate the dressed effective Coulomb potential of the monolayer phosphorene encapsulated by polar hBN slabs and briefly describe the theory and formalism of linear response function, random phase approximation and collective mode extraction procedure, define our two models under consideration and also figure out the semiclassical framework for anisotropic systems. In Sec.~\ref{sec3} we present and describe main results of the collective modes along different directions of the system and finally, we conclude and summarize our main results in Sec.~\ref{sec4}.
%
\section{Model and Theory}
\label{sec2}
\subsection{Model Hamiltonian}
\label{sec2a}
%
We consider a clean doped monolayer phosphorene at finite temperature. The electronic band structure of phosphorene has been calculated based on VASP and SIESTA density-functional theory packages \cite{Elahi}. The calculations based on both packages provide identical trends for the first conduction band in the vicinity of $\Gamma$-point, which is the exact position of the Conduction Band Minimum (CBM). For the first valence band, however, the actual position of the Valence Band Maximum (VBM) slightly differs from SIESTA to VASP. While the SIESTA band structure predicts the VBM to be located precisely at $\Gamma$-point, the VASP package suggests an indirect band gap with its actual valence maximum occurring along the $\Gamma-Y$ high-symmetry direction. The discrepancy can be attributed to the different calculation methods employed, i.e. the pseudopotential scheme combined with atomic orbitals in SIESTA versus projector augmented-wave formalism with plane waves in VASP. Accordingly a very marginal change of the overlap between atomic orbitals would transform the nature of the band gap from direct to indirect.

Having ignored this slightly discrepancy of the position of the VBM, we can assume that the system has a direct band gap. With this, we could write down a low-energy model Hamiltonian \cite{Zare}. Basically, the electronic band structure could be described by a four band model in the tight-binding model, however, it can be expressed by a two-band model owing to the C$_{2h}$ point group invariance. Expanding the tight-binding \cite{Ezawa,Katsnelson,Rodin} model around the $\Gamma$ point, one obtains the low-energy ${\bf k} \cdot {\bf p}$ model for phosphorene \cite{Zare} as,
\begin{equation}\label{Hamiltonian}
H_{eff}= \left(
\begin{array}{cc}
	E_c +  \eta_c k_x^2 + \nu_c k_y^2 & \quad
	\gamma k_x \\
	\gamma k_x & \quad
	E_v - \eta_v k_x^2 - \nu_v k_y^2
\end{array}
\right)
\end{equation}
in the conduction and valence band basis. $E_\tau$ with $\tau$ denoting the conduction or valence band, is the band edge at the $\Gamma$ point with direct energy gap $E_g = E_c - E_v$ and the off-diagonal $\gamma k_x$ element with the real parameter $\gamma$ is the interband coupling term. Other parameters can be extracted from the knowledge of DFT results \cite{Elahi} where we have $E_g=0.912$ eV and the effective masses $m_{{c,x}}=0.146$, $m_{{v,x}}=0.131$, $m_{{c,y}}=1.240$ and $m_{{v,y}}=7.857$ in units of the electron bare mass $m_0$, which implies that $\gamma=0.480$ $\mathrm{eV \cdot nm}$, $\eta_c=0.008$, $\eta_v=0.038$ in units of $\mathrm{eV \cdot nm^{2}}$, $\nu_c=0.030$ and $\nu_v=0.005$ in units of $\mathrm{eV \cdot nm^2}$. Notice that the hole effective mass along the zigzag direction ($k_y$-axis) is almost $7$ times greater than that along the armchair direction ($k_x$-axis) which induces strong in-plane anisotropy. For definition of the armchair and zigzag directions, see Figs. \ref{fig1}(a) and \ref{fig1}(b). This accurate low-energy model Hamiltonian protected all needed symmetries. As discussed in \cite{Zare}, this model Hamiltonian provides the correct Fermi surface in comparison with those proposed in \cite{Rodin} and \cite{Low}, because the former suggests entirely inappropriate effective mass values of charge carriers in phosphorene, while the latter breaks the time-reversal symmetry even at low electron or hole density.

In the 2D phosphorene case, the isofrequency profiles are obtained by horizontally cutting the dispersion surface separately. We illustrate an isofrequency contour surface in the $k$-space to explore their symmetries for ${\varepsilon^\tau_{\bf{k}_{\rm F}}} = 0.05-0.5$ eV with step of $0.05$ eV in both the electron and hole doped cases in Fig. \ref{fig1}. The (c) and (d) panels refer to the Fermi surfaces for the electron and hole doped systems, respectively. As it can be inferred from the Fig. \ref{fig1}, the isofrequency contour Fermi boundaries in the electron-doped case are elliptic-like with a good precision. However, by elevating of the Fermi energy in the hole-doped case, the Fermi boundaries show a deviation from elliptical form to a boxlike form in the $k_y$ direction and even concave boundaries emerge out which result in the Von Hove singularities in the corresponding direction. These features stem from the fact that the quasiparticles (especially the holes) along the zigzag direction are effectively much heavier than that of the armchair direction. Moreover, it can be inferred that the off-diagonal element $\gamma k_x$ of the Hamiltonian (\ref{Hamiltonian}) as the interband coupling term has a significant role in the physics of the Fermi boundaries \cite{Zare}. We set the zero energy at the conduction band minimum edge in the electron doped and at the valence band maximum edge in the hole doped cases, respectively.
\vspace*{-0.4cm}
\subsection{Theory and Formalism}
\label{sec2b}
%
The dielectric constant of hBN has been described by a tensor of rank 2, and we can reduce the number of independent optical constant tensor elements by considering its lattice symmetries. In hBN lattice owing to the rotational symmetry, together with the reflection and inversion operators, the optical constant matrix consists of only three diagonal terms where the dielectric tensor casts in the form diag$[\epsilon_x,\epsilon_y,\epsilon_z]$ in which $\epsilon_x = \epsilon_y \equiv \epsilon_{\perp}$ in a Cartesian frame of reference oriented along the principal axes of the crystal. $\epsilon_{i}$ in general depends on the angular frequency $\omega$. The choice of $\epsilon_{\perp} > 0$ where $\epsilon_z <0$ corresponds to a twofold hyperbolic and choice of $\epsilon_{\perp}< 0$ where $\epsilon_z > 0$ describes a one-fold hyperbolic. Note that the unique feature of a hyperbolic environment is the strongly directional light emission.

\begin{table}
\centering
\caption{The microscopic parameters related to hBN dielectric tensor components in Eq. (\ref{ep}) extracted from \cite{Caldwell,Tomadin}.}
\begin{tabular}{cccccc}
  \hline\noalign{\smallskip}
  $l$ & $\epsilon_{l,0}$ & $\epsilon_{l,\infty}$ & $\gamma_l (meV)$ & $\hbar \omega_l^T (meV)$  & $\hbar \omega_l^L (meV)$ \\
  \noalign{\smallskip}\hline\noalign{\smallskip}
  $x$ & 6.41 & 4.54 & 0.82 & 168.0 & 199.6 \\
  $z$ & 3.00 & 2.50 & 0.23 & \, 94.2 & 103.2 \\
  \noalign{\smallskip}\hline
\end{tabular}
\label{tab1}
\end{table}
In our study, we consider a vertical heterostructure composed of a phosphorene sheet located at $z=0$ and encapsulated with two homogenous anisotropic insulators of uniaxial hBN slabs with thicknesses $d \sim {60}$ nm and $d' {\sim} 10$ nm with optical onset depending on the angular frequency. This sandwich is considered to be deposited on a semi-infinite SiO$_2$ substrate and the whole resulted heterostructure is embedded in air (see Fig. \ref{fig2}). The electric potential of the system can be calculated from the following elementary approach. The displacement field ${\bf D}({\bf r},z)$ must satisfy the condition $\nabla \cdot {\bf D}({\bf r},z) = 0$ everywhere in space. However, the presence of an electron with charge density $-e\delta^2({\bf r})\delta(z)$ at $z=0$ implies a discontinuity in the normal component of the displacement field across $z=0$ plane. Imposing the boundary conditions on the Laplace equations which are given by 
\begin{equation}\label{bc}
  \begin{cases}\vspace{0.1cm}
    -\epsilon_3 \big[\partial^2_{\bf r} \phi_3({\bf r},z) ~  + ~ \partial^2_z\phi_3({\bf r},z) \big] \;\;\, = 0, &  z>d'\\ \vspace{0.1cm}
    -\epsilon_x\partial^2_{x}\phi_{2B}({\bf r},z)-\epsilon_y\partial^2_{y}\phi_{2B}({\bf r},z) \\ \vspace{0.1cm} ~~~~~~~~~~~~~~~~~~~~ -\epsilon_z\partial^2_z\phi_{2B}({\bf r},z) = 0, &  0<z<d'\\ \vspace{0.1cm}
    -\epsilon_x\partial^2_{x}\phi_{2A}({\bf r},z)-\epsilon_y\partial^2_{y}\phi_{2A}({\bf r},z) \\ \vspace{0.1cm} ~~~~~~~~~~~~~~~~~~~~ -\epsilon_z\partial^2_z\phi_{2A}({\bf r},z) = 0, & -d<z<0\\ \vspace{0.1cm}
    -\epsilon_1\big[ \partial^2_{\bf r} \phi_1({\bf r},z) ~ + ~ \partial^2_z\phi_1({\bf r},z) \big] \;\;\, = 0, &  z<-d
  \end{cases}
\end{equation}
in which $\epsilon_1$ and $\epsilon_3$ respectively refer to the dielectric constants of the environments below and over the system, together with the Laplace equations, lead to the 2D Fourier transform of the dressed potential $V(q,\omega)$ as \cite{Lundeberg} 
\begin{widetext}
\begin{eqnarray}
\label{Vqw}
\nonumber
V(q,\omega) = v({q})
&\Bigg[&
\sqrt{\epsilon_x\epsilon_z} +
\frac{\epsilon_3\epsilon_1}{\sqrt{\epsilon_x\epsilon_z}} +
\Bigg(\sqrt{\epsilon_x\epsilon_z}-\frac{\epsilon_3\epsilon_1}{\sqrt{\epsilon_x\epsilon_z}}\Bigg)
\frac{\cosh\Big(q\sqrt{\frac{\epsilon_x}{\epsilon_z}}(d'-d)\Big)}
     {\cosh\Big(q\sqrt{\frac{\epsilon_x}{\epsilon_z}}(d'+d)\Big)} +
(\epsilon_3 - \epsilon_1)
\frac{\sinh\Big(q\sqrt{\frac{\epsilon_x}{\epsilon_z}}(d'-d)\Big)}
     {\cosh\Big(q\sqrt{\frac{\epsilon_x}{\epsilon_z}}(d'+d)\Big)}\\
&~& ~ + (\epsilon_3 + \epsilon_1)
\tanh\Bigg(q\sqrt{\frac{\epsilon_x}{\epsilon_z}}(d'+d)\Bigg)
\Bigg]
\Bigg[
(\epsilon_3 + \epsilon_1)
\sqrt{\epsilon_x\epsilon_z}
+ (\epsilon_3\epsilon_1 + \epsilon_x\epsilon_z)
\tanh\Bigg(q\sqrt{\frac{\epsilon_x}{\epsilon_z}}(d'+d)\Bigg)
\Bigg]^{-1}
\end{eqnarray}
\end{widetext}
where $v(q)=2\pi e^2/q$ is the Fourier transform of the bare Coulomb potential in 2D electron systems. The frequency dependence of the dressed Coulomb interaction is owing to the optical phonons in the hBN slabs. The components of the uniaxial dielectric tensor of hBN have an important dependency on frequency in the mid-infrared range, which is parameterized in the following form
\begin{equation}\label{ep}
\epsilon_l(\omega)=\epsilon_{l,\infty}+\frac{\epsilon_{l,0}-\epsilon_{l,\infty}}{1-(\omega/\omega_l^T)^2 - i\gamma_l\hbar\omega/(\hbar\omega_l^T)^2}
\end{equation}
with $l=x$ or $z$ and $\epsilon_{l,0}$ and $\epsilon_{l,\infty}$ are the static and the high-frequency dielectric constants, respectively, while $\omega_l^T$ and $\gamma_l$ are the transverse optical phonon frequency and it corresponding amplitude decay rate in the $l$-direction. The parameters in Eq. (\ref{ep}) are given in Refs. \cite{Caldwell,Tomadin} and also reported in Table~\ref{tab1}.

In order to consider the impact of the many-body effects, we define the dynamic dielectric function within the Random Phase Approximation (RPA) as 
\begin{equation}\label{RPA}
\epsilon({\bf q},\omega) = 1 -V({\bf q},\omega) \chi_0 ({\bf q},\omega)
\end{equation}
where $\chi_0 ({\bf q},\omega)$ is the noninteracting density-density linear response function (or the Lindhard function) \cite{Giuliani} 
\begin{eqnarray}\label{Lindhard} 
&& \chi_0(\textbf{q},\omega) = \nonumber\\
&&~~-\frac{g_s}{(2\pi)^2} \int d\textbf{k}
\sum_{\tau,\tau'} \frac{ f (\varepsilon^\tau_\textbf{k}) - f(\varepsilon^{\tau'}_{\textbf{k}'})    }{\varepsilon^\tau_\textbf{k} - \varepsilon^{\tau'}_{\textbf{k}'} + \hbar \omega + \imath \eta} \vert \langle \psi^\tau_\textbf{k} | \psi^{\tau'}_{\textbf{k}'} \rangle \vert^2 \quad
\end{eqnarray}
in which $\textbf{k}' \equiv \textbf{k}+\textbf{q}$, the spin degeneracy for the phosphorene is $g_s=2$, the Fermi-Dirac distribution function is $f (\varepsilon) = \{\exp{ [ (\varepsilon - \mu)/k_B T ]}+1\}^{-1}$ and the chemical potential $\mu$ is obtained through the charge carrier concentration of the system, $n=\int_{-\infty}^{\mu}D(\varepsilon)f(\varepsilon)d\varepsilon$. Again, $\tau$ refers to the conduction or valence band and $\eta \rightarrow 0^+$ is the fingerprint of the causality of the retarded response function due to the adiabatic principle and has also a phenomenological broadening nature. In this equation, $\varepsilon^\tau_\textbf{k}$ and $\psi^\tau_\textbf{k}$ are the eigenvalue and eigenvector of the Hamiltonian, respectively. It is worth mentioning that it is absolutely necessary that $\langle \psi^\tau_\textbf{k} | \psi^{\tau'}_{\textbf{k}'} \rangle$, which is no longer a unity, to be considered in the calculation. 

According to Eq. (\ref{RPA}), the RPA dielectric function is a complex valued function $\epsilon(\textbf{q},\omega)=\epsilon_1(\textbf{q},\omega)+ \imath \, \epsilon_2(\textbf{q},\omega)$. Notice that the real part of the dielectric function $\epsilon_1$ describes the reflection of light, while the imaginary part $\epsilon_2$ describes the absorption of light (inelastic process). This holds for any desired model including our both simple and coupled models defined in the following. Regardless of the applied model, collective modes can be extracted from the poles of the RPA dielectric function. Accordingly, for each $\textbf{q}$ both the real part $\epsilon_1$ and the imaginary part of the dielectric function $\epsilon_2$ should simultaneously vanish. The imaginary part in the intraband single-particle excitation region dubbed as the electron-hole continuum and defined by the upper/lower boundaries $\hbar \omega^\tau_\pm (\textbf{q}) = \varepsilon^\tau_{\textbf{q} \pm \textbf{k}_F} - \varepsilon^\tau_{\textbf{k}_F}$ has nonzero value (Landau damping) owing to the intraband processes and vanishes outside. Therefore, finding the poles of $\epsilon$ reduces to finding the roots of its real part $\epsilon_1$ above the upper boundary of the continuum $\hbar \omega^\tau_+(\textbf{q})$. Moreover, it will be shown later in this paper that the two Reststrahlen bands inherited from hBN with energy intervals $[\hbar \omega_z^T , \hbar \omega_z^L ]$ and $[\hbar \omega_x^T , \hbar \omega_x^L ]$ possess some localized Surface Plasmon-Phonon-Polariton (SPPP) modes which are also obtained by the same procedure of finding the roots of the $\epsilon_1$ above the electron-hole continuum for each $\textbf{q}$. This will happen in the coupled model described below.

Finally, it is worthwhile to mention that for any plasmon mode with dispersion $\hbar \omega_p(\textbf{q})$ at wavevector $\textbf{q}$, it is straightforward to calculate the damping parameter 
\begin{equation}\label{damping}
\gamma (\textbf{q}, \omega_p(\textbf{q})) =
\frac{\Im m \chi_0 (\textbf{q}, \omega_p(\textbf{q}))}{\partial  \Re e \chi_0 (\textbf{q}, \omega)  / \partial \omega |_{\omega = \omega_p(\textbf{q})}}
\end{equation}
whose rescaled version $\gamma (\textbf{q}, \omega_p(\textbf{q})) / \omega_p(\textbf{q})$ in units of the plasmon angular frequency is a powerful indicator of the damping intensity of the collective mode at that wavevector.

%
\subsection{Simple and Coupled Models}
\label{sec2c}
%
Two separate models are investigated in order to perceive the plasmon modes in the structure. First of all, we ignore both the thin film effects and the polar modes of the two hBN slabs encapsulated the monolayer phosphorene; a model which is called \textit{the simple model}. Thus, the bare Coulomb potential $v(q)=4\pi e^2/(\epsilon_1+\epsilon_3) \, q$ is experienced by the charge carriers of phosphorene. In this simple model $d$ and $d'$ both go to infinity and so the two half-space media are both hBN with $\epsilon_1=\epsilon_3=4.5$. In this simple model, the plasmon mode of the phosphorene as a gapless excitation also emerges out. 

In the second case, which is called \textit{the coupled model}, we consider both the thin film effects of the hBN slabs encapsulated the monolayer phosphorene and the electron-phonon coupling of the phosphorene's charge carriers with the damping phonon-polaritons of the hBN slabs. Therefore, the dressed Coulomb potential yielded in Eq. (\ref{Vqw}) is experienced by the phosphorene's charge carriers and the two half-space media are the SiO$_2$ substrate with $\epsilon_1=3.9$ in below and vacuum with $\epsilon_3=1.0$ in above (see Fig. \ref{fig2}). In the coupled model, plasmons would couple to the phonon-polaritons of the hBN slabs and result in trifurcated SPPP modes.
%
\subsection{Semiclassical Model}
\label{sec2d}
%
Moreover, aiming to have a better understanding of the long-wavelength limit $q \ll k_F$ of the collective modes' behavior and their losses due to the absorption and the intraband processes, it is fruitful to compare and contrast our results based on the RPA approximation with that of the semiclassical (SC) model based on the Maxwells' theory of classical electrodynamics. A phosphorene sheet sandwiched between two media $i=1,2$ (from below and above, respectively) with dielectric functions $\epsilon_i$ and the averaged value $\bar{\epsilon}=(\epsilon_1+\epsilon_2)/2$ is considered. Since we are exclusively interested in the intraband processes a single-band model suffices and thus the band indicator in the following formulas will be omitted. Here we arbitrarily focus on the valence band of a hole-doped system with the hole density $n$ and with the hole effective masses $m_j=m_{v,j}$ along the armchair and the zigzag directions $j=x,y$ and with $g_{2D}=m_d/\pi \hbar^2$ and $m_d=\sqrt{m_x m_y}$ respectively as the elliptic 2D density-of-states at the Fermi level and its corresponding average charge carrier effective mass.

In our calculations, the relaxation time $\tau$ between two successive scattering of an electron from the lattice is $\tau=\hbar/\eta$ with the aforesaid $\eta\rightarrow 0^+$. Consequently the longitudinal optical conductivity $\sigma_{jj}(\omega)$ along the armchair and zigzag directions according to the naive Durde theory are \cite{Ashcroft}
\begin{equation}\label{conductivityDef}
\sigma_{jj} (\omega) = \frac{n e^2 \tau}{m_j \left( 1- \imath \omega \tau \right)}
\end{equation}
that can be generalized along an arbitrary direction making angle $\theta$ with the $k_x$-axis and in both of the $\beta=s,p$ modes as a tensor $\mathbf{M}(\theta)$ with elements $M_{\beta\beta'}(\theta)$: 
\begin{eqnarray}\nonumber
&& \mathbf{M}(\theta) \equiv
\left(
\begin{array}{cc}\vspace{0.2cm}
 M_{ss}(\theta) & M_{sp}(\theta)  \\
 M_{ps}(\theta) & M_{pp}(\theta) 
\end{array}
\right) \quad\quad
\\ \label{MDef}
&& \; = 
\left(
\begin{array}{cc}\vspace{0.2cm}
 \sigma_{xx}\sin^2\theta+\sigma_{yy}\cos^2\theta & \left(\sigma_{yy}-\sigma_{xx}\right)\sin\theta\cos\theta  \\
 \left(\sigma_{yy}-\sigma_{xx}\right)\sin\theta\cos\theta & \sigma_{xx}\cos^2\theta+\sigma_{yy}\sin^2\theta
\end{array}
\right)\quad\quad
\end{eqnarray}
which its determinant $|{\mathbf{M}(\theta)}|=\sigma_{xx}\sigma_{yy}$ is independent of direction. On the other hand, the total admittance of each of the $\beta=s,p$ modes is the summation of the two media's corresponding admittances $Y_\beta=Y_\beta^1+Y_\beta^2$ \cite{Low} resulting in the diagonal tensor  
\begin{equation}\label{Y}
\mathbf{Y}\equiv
\left(\begin{array}{cc}
 Y_s & 0  \\
 0   & Y_p    
\end{array}\right)
=
\left(\begin{array}{cc}
\frac{\imath\left( \kappa_1 + \kappa_2 \right)}{\omega \epsilon_0} & 0 \\
0 & -\imath \omega \epsilon_0 \left( \frac{\epsilon_1}{\kappa_1} + \frac{\epsilon_2}{\kappa_2}  \right)
\end{array}\right)
\end{equation}
in which the following definition is assumed:
\begin{equation}\label{ki}
\kappa_{i}^2 \equiv q^2 - \epsilon_i \; \omega^2 / c^2
\end{equation}
Now the boundary condition of the transverse magnetic (TM) or the p-polarized electromagnetic wave at $z=0$ interface (as the continuity and a jump respectively in the electric and magnetic field's tangential components) demand the determinant of the summation of $\mathbf{Y}$ and $\mathbf{M}(\theta)$ to vanish $\left| \mathbf{Y} + \mathbf{M}(\theta) \right| = 0$: 
\begin{equation}\label{LowDispersion}
\left( Y_s + M_{ss} \right) \left( Y_p + M_{pp} \right) - M_{sp} M_{ps}= 0
\end{equation}
which is identical to the dispersion mentioned in Ref. \cite{Low}. It can be written in a more explicit form:
\begin{equation}\label{fundamental}
Y_s Y_p + \sigma_{xx}\sigma_{yy} + \left( Y_s M_{pp}(\theta) + Y_p M_{ss}(\theta) \right) = 0
\end{equation}
This is the long-wavelength limit of the collective mode's semiclassical dispersion relation generalized for an anisotropic system. Furthermore, in the nonretarded regime $q\gg\sqrt{\epsilon_i} \omega/c$, Eq. (\ref{ki}) reduces to $\kappa\equiv \kappa_i= q$ and Eq. (\ref{Y}) results in $Y_s = 2\imath q/\omega \epsilon_0$, $Y_p =2 \omega \epsilon_0\bar{\epsilon}/\imath q$ and $Y_s Y_p = 4 \bar{\epsilon}$. Thus the term $Y_p M_{ss}$ in Eq. (\ref{fundamental}) in comparison with $Y_s M_{pp}$ is ignorable and consequently the whole dispersion solved for $\kappa$ can be simplified: 
\begin{equation}\label{q_implicit}
\kappa = \frac{\imath\omega \epsilon_0}{2}
\frac{4 \bar{\epsilon} + \sigma_{xx}\sigma_{yy}}{\sigma_{xx}\cos^2{\theta}+\sigma_{yy}\sin^2{\theta}}
\end{equation}
It should be emphasized that the dispersion Eq. (\ref{q_implicit}) has the additional term $\sigma_{xx}\sigma_{yy}$ in its nominator compared with its equivalent equation in the Ref. \cite{Low}. But as it will become obvious at the end, it is in the second order of the fine-structure constant $\alpha=e^2/4\pi\epsilon_0 \hbar c$ and can be ignored, but it is fruitful to keep it until the final stages. By defining the $p$-mode directional mass as
\begin{equation}\label{directionalMass}
m_p(\theta)\equiv \frac{m_x m_y}{ m_y \cos^2 \theta + m_x \sin^2 \theta }
\end{equation}
and its scaled version $\bar{m}_p(\theta)\equiv m_p(\theta)/m_d$ in units of $m_d$ and then insertion of Eq. (\ref{conductivityDef}), explicit calculation of Eq. (\ref{q_implicit}) is straightforward: 
\begin{eqnarray}
\nonumber
\kappa &=& 
\frac{2 \bar{\epsilon} \epsilon_0 m_p(\theta)}{n e^2} \omega^2
- \frac{\epsilon_0 ne^2 m_p(\theta)}{2 m^2_d} 
\frac{\left(\omega \tau \right)^2}{1 + \left(\omega \tau\right)^2}
 \\
 \label{kappa_total}
&+& \imath 
\left(
\frac{2 \bar{\epsilon} \epsilon_0 m_p(\theta)}{n e^2} \frac{\omega}{\tau}
+ \frac{\epsilon_0 ne^2 m_p(\theta)}{2 m^2_d}
\frac{\omega \tau }{1 + \left(\omega \tau\right)^2}
\right)\quad
\end{eqnarray}
The real part of $\kappa$ indicates the confinement degree of the plasmon's electromagnetic field within the dielectric media and its reciprocal value is the \emph{penetration depth} of the surface plasmon $\zeta_{p} \equiv 1/ \Re e \left\lbrace \kappa \right\rbrace$ \cite{Goncalves}: 
\begin{equation}
\label{penetrationDepth}
\zeta_{p} 
=
\frac{1}{\epsilon_0 m_p(\theta)}
 \left[ \frac{2 \bar{\epsilon}\omega^2 }{ne^2} 
 - \frac{n e^2}{2 m^2_d} \frac{(\omega \tau)^2}{1+(\omega \tau)^2} \right]^{-1}
\end{equation}
Above the overdamping regime $\omega \tau \gg 1$ (equivalently $\hbar \omega\gg \eta$) the second term of the real part of $\kappa$ in the plasmon dispersion Eq. (\ref{kappa_total}) tends to $-\epsilon_0 e^2 m_p / 2 m_d^2$ which is ignorable compared to the first term and the second term in the imaginary part vanishes explicitly: 
\begin{equation}\label{answer}
\kappa =  
\frac{\bar{\epsilon}}{2 \alpha}
\frac{m_p(\theta)}{m_d}
\frac{\hbar \omega}{n / g_{2D}}
\left(
\frac{\omega}{c} 
+ \imath 
\frac{\eta}{\hbar c}
\right)
\end{equation}

Considering the wavenumber as a complex quantity $\kappa=q=q'+ \imath q''$, the \emph{propagation length} of the surface plasmon $L_{p} \equiv 1 / 2 \Im m\left\lbrace q \right\rbrace = 1 / 2 q''$ defined as the distance by which its intensity suppresses to $1/e$ of its initial value and also the \emph{wavelength} of the surface plasmon $\lambda_{p} \equiv 2 \pi / \Re e \left\lbrace q \right\rbrace = 2 \pi / q'$ relative to its (laser) stimulator light wavelength in vacuum $\lambda_\omega=2 \pi c / \omega$ can be retrieved. Insertion of the average dielectric constant of the phosphorene sheet's surrounding media $\bar{\epsilon}$ into Eq. (\ref{answer}) and in latter definitions yield the final results.

In the simple model the two surrounding media's dielectric constants are equal $\bar{\epsilon}_s=\epsilon_i=4.5$ resulting in: 
\begin{eqnarray}
\label{Lsp}
L_{p}
&\equiv&
\frac{ 1}{2 \Im m\left\lbrace q \right\rbrace}
=
\frac{\alpha}{\bar{\epsilon}_s}
\frac{m_d}{m_p(\theta)}
\frac{\hbar c}{\eta} 
\frac{n/g_{2D}}{\hbar \omega}
\\
\label{lambdaP0}
\frac{\lambda_{p}}{\lambda_\omega}
&\equiv&
\frac{\omega}{c}
\frac{1}{\Re e\left\lbrace q \right\rbrace}
=
\frac{2 \alpha}{\bar{\epsilon}_s}
\frac{m_d}{m_p(\theta)}
\frac{n/g_{2D}}{\hbar \omega}
\end{eqnarray}
which are the anisotropic generalization of the well-known answers for graphene via the presence of the $m_d/m_p(\theta)$ coefficient. As a rule of thumb embracing the elliptic Fermi contour approximation $n=g_{2D} E_F$ yields: 
\begin{equation}\label{RuleofThumb}
L_{p} = \frac{\alpha}{\bar{\epsilon}_s} \frac{m_d}{m_p(\theta)} \frac{\hbar c}{\eta} \frac{E_F}{\hbar \omega}, \;\;
\lambda_{p}/\lambda_\omega = \frac{2 \alpha}{\bar{\epsilon}_s} \frac{m_d}{m_p(\theta)} \frac{E_F}{\hbar \omega}
\end{equation}

The surrounding media's average dielectric constant $\bar{\epsilon}_c=2.45$ in the coupled model is now multiplied by the effective dielectric function $\epsilon_{eff}(q,\omega) = v(q) / V(q,\omega)$ of the heterostructure applied at $z=0$ implicitly expressed in Eq. (\ref{Vqw}). This results in $\bar{\epsilon} (q,\omega)= \epsilon_{eff}(q,\omega) \, \bar{\epsilon}_c$. So, the dispersion Eq. (\ref{answer}) for the complex wavevector $q(\theta, \omega)=\kappa$ casts in a self-consistent form:
\begin{equation}\label{answerC}
q_{i+1} = \epsilon_{eff} \left( q_i, \omega \right) \frac{\bar{\epsilon}_c}{2 \alpha} \frac{m_p(\theta)}{m_d} \frac{\hbar \omega}{n / g_{2D}} \left( \frac{\omega}{c} + \imath \frac{\eta}{\hbar c}\right)
\end{equation}
with the initial value $q_0=0$. In our calculations this sequence converges quickly up to $i=5$. Consequently evaluation of the dispersion of SPPPs $p_i$ and their before-mentioned related quantities such as $L_{p_i}$ and $\lambda_{p_i}/\lambda_\omega$ are viable.  
\vspace*{-0.1cm}
\section{Numerical Results and Discussion}
\label{sec3}

%
\begin{figure}
	\centering
	\includegraphics[width=\linewidth]{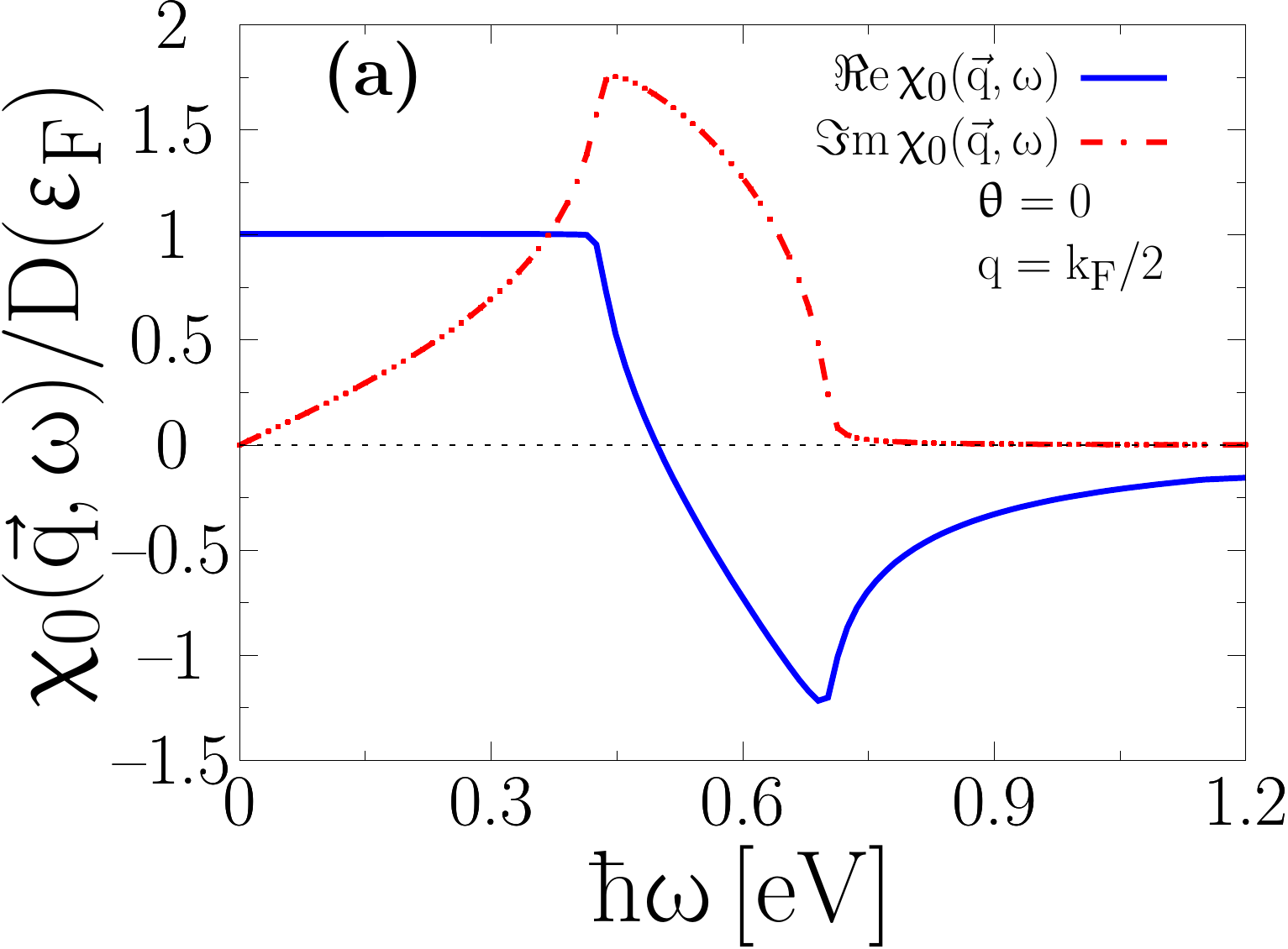}
	\includegraphics[width=\linewidth]{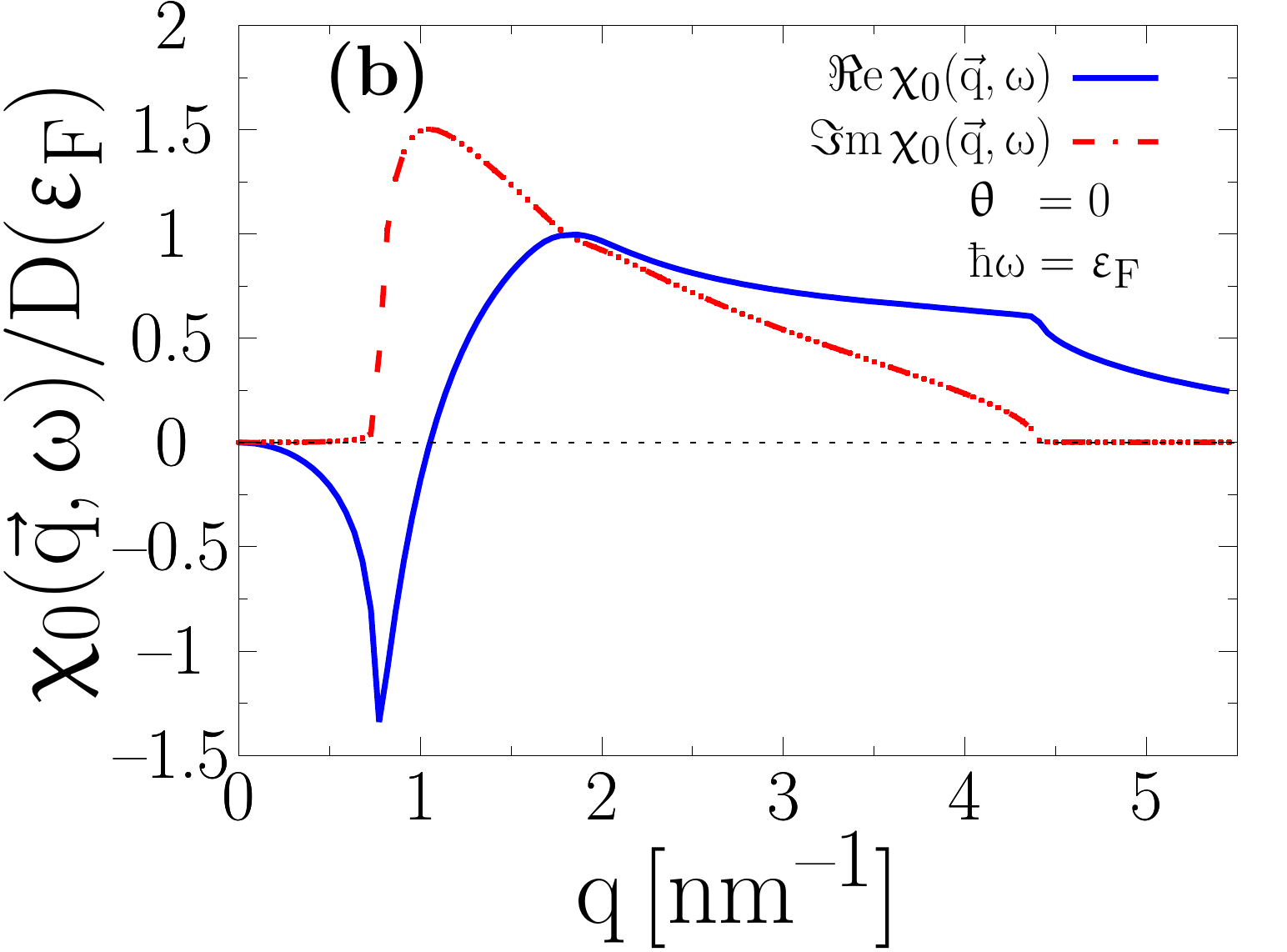}
	\caption{(Color online) Noninteracting density-density linear response function $\mathrm{\chi_0(\textbf{q},\omega)}$ in units of density-of-states at Fermi energy D$(\mathrm{\varepsilon_F})$ as a function of (a) $\mathrm{\omega}$ for $\mathrm{q=k_F (\theta = 0)/2}$ and (b) $\mathrm{q}$ for $\mathrm{\hbar\omega=\varepsilon_F}$ along the armchair direction in the electron doped case at zero temperature. The real and imaginary parts are illustrated by blue solid line and red dotted-dashed line respectively.}
	\label{fig3}
\end{figure}
\begin{figure}
	\centering
	\includegraphics[width=\linewidth]{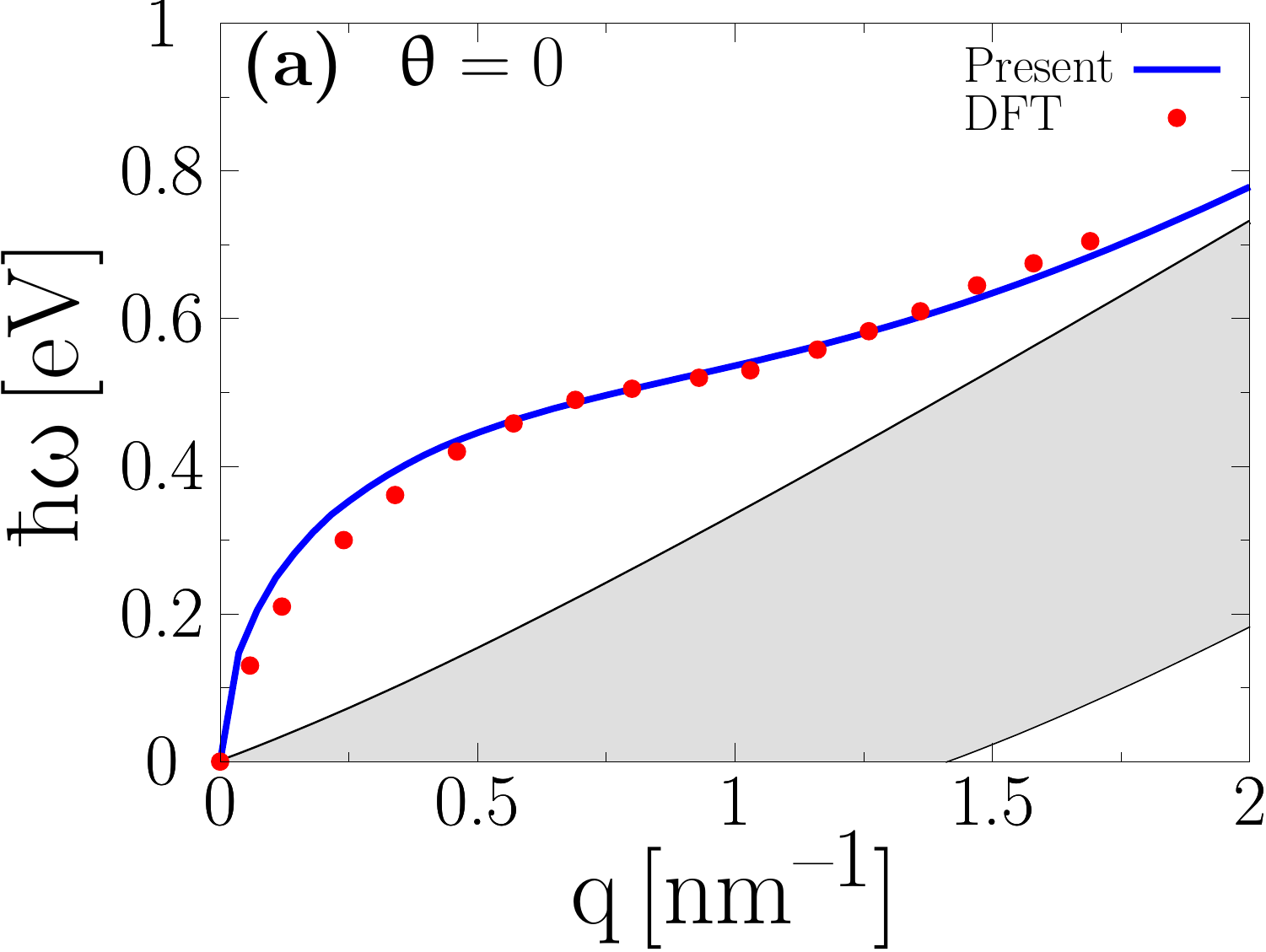}
	\includegraphics[width=\linewidth]{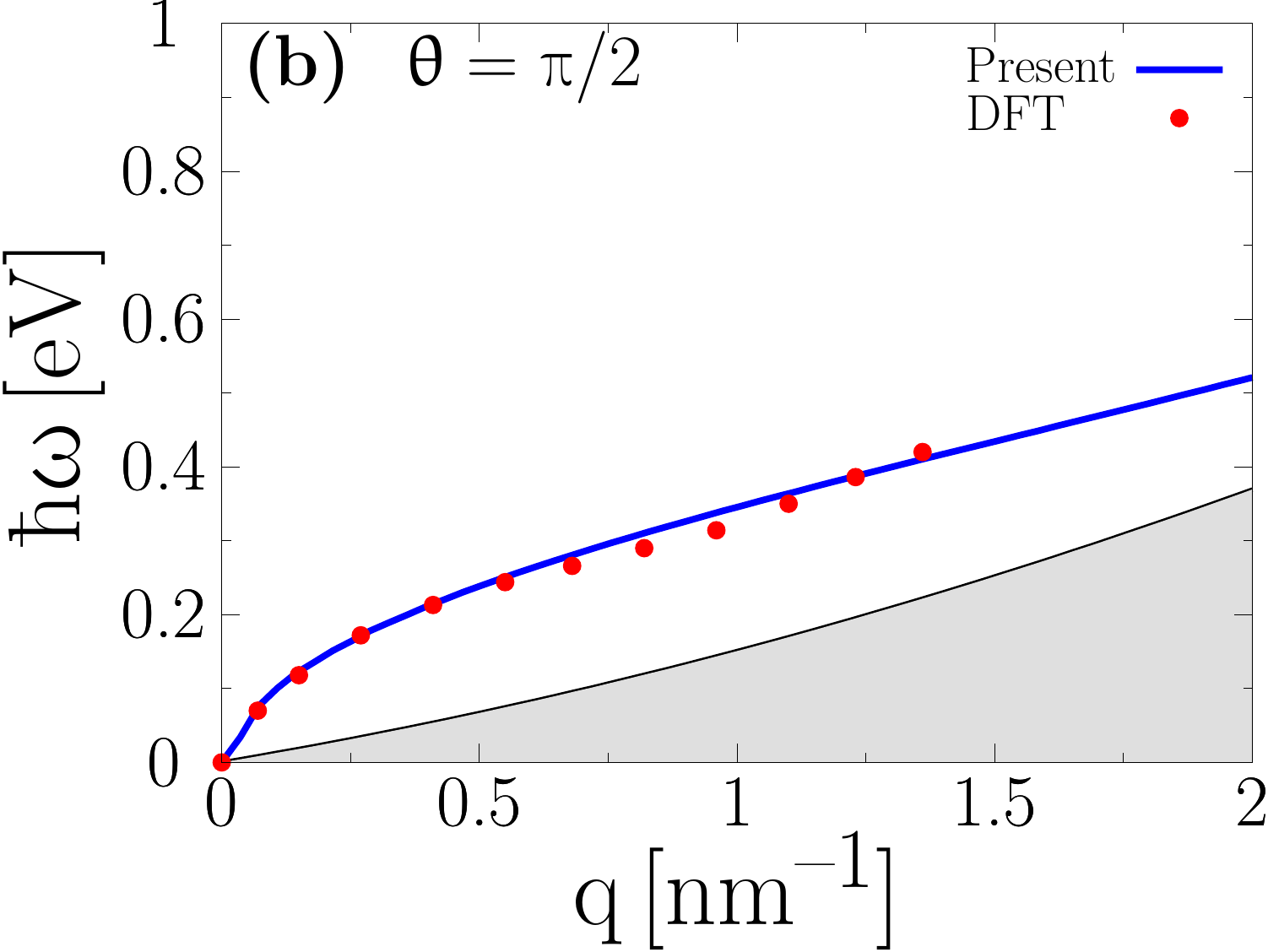}
	\caption{(Color online) The plasmon mode $\mathrm{\hbar \omega_p(\textbf{q})}$ of the free-standing phosphorene as a function of wavenumber $\mathrm{q}$ - the solid blue lines - along the (a) armchair ($\mathrm{\theta=0}$) and (b) zigzag ($\mathrm{\theta=\pi/2}$) directions in comparison with those results obtained by DFT-RPA approximation \cite{TorbatianPRB} - the red dots - in zero temperature and in the case that the dielectric constants of the environment are unity (the phosphorene is embedded in air) and the electron density is $\mathrm{n_e=3.5\times 10^{13} \; cm^{-2}}$. The gray region is the intraband electron-hole continuum. At the long-wavelength limit the plasmon mode behaves like $\mathrm{\sqrt{q}}$.}
	\label{fig4}
\end{figure}
In this section, our numerical results for the plasmon-phonon-polaritons in an encapsulated phosphorene are presented. We investigate the plasmon modes and the SPPP modes in the system and explore the impact of the hybrid material which creates three new plasmon-phonon-polariton modes with dispersion relations that are distinctly different from the original phosphorene plasmon dispersion. They all exhibit a drastic anisotropy inherited from the band structure described by the model Hamiltonian Eq. (\ref{Hamiltonian}).

In the first step, we calculate the noninteracting density-density linear response function $\chi_0(\textbf{q},\omega)$ defined in Eq. (\ref{Lindhard}) for which the value $\eta = 10^{-3} \times |\mu|$ has been used all over this study. The real and imaginary parts of the response function (in units of $D(\varepsilon_F)$, the density-of-states at Fermi energy) along the $k_x$-axis are plotted in Fig. \ref{fig3}. For the sake of simplification, we use zero temperature in this figure. In Fig. \ref{fig3}(a), $\chi_0(\textbf{q},\omega)$ as a function of frequency $\omega$ for $q_x=k_F (\theta = 0)/2$ is presented. The real part $\Re e \chi_0(\textbf{q},\omega)$ starts from unity and changes sign from positive to negative as $\omega$ sweeps across the electron-hole continuum, while the imaginary part $\Im m \chi_0(\textbf{q},\omega)$ shows a sharp cutoff. The Lindhard function $\chi_0(\textbf{q},\omega)$ versus momentum $q$ for $\hbar\omega=\varepsilon_{\rm F}$ is also illustrated in Fig. \ref{fig3}(b). The real part $\Re e \chi_0(\textbf{q},\omega)$ shows two different cusps when $q$ attains to the electron-hole continuum.
\begin{figure}
	\hspace*{-0.2cm}
	\includegraphics[width=0.95\linewidth]{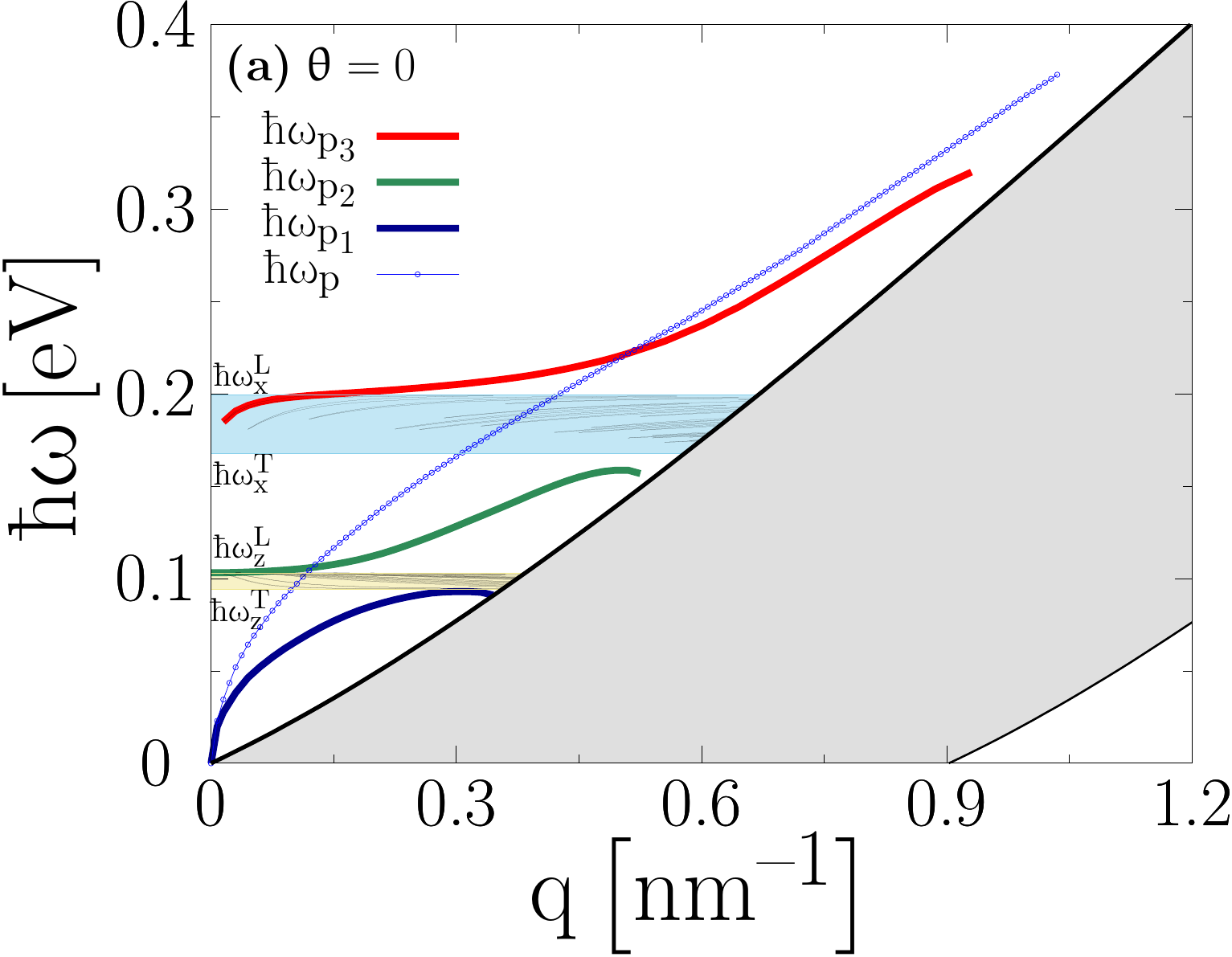}\\
	\hspace*{-0.2cm}
	\includegraphics[width=0.95\linewidth]{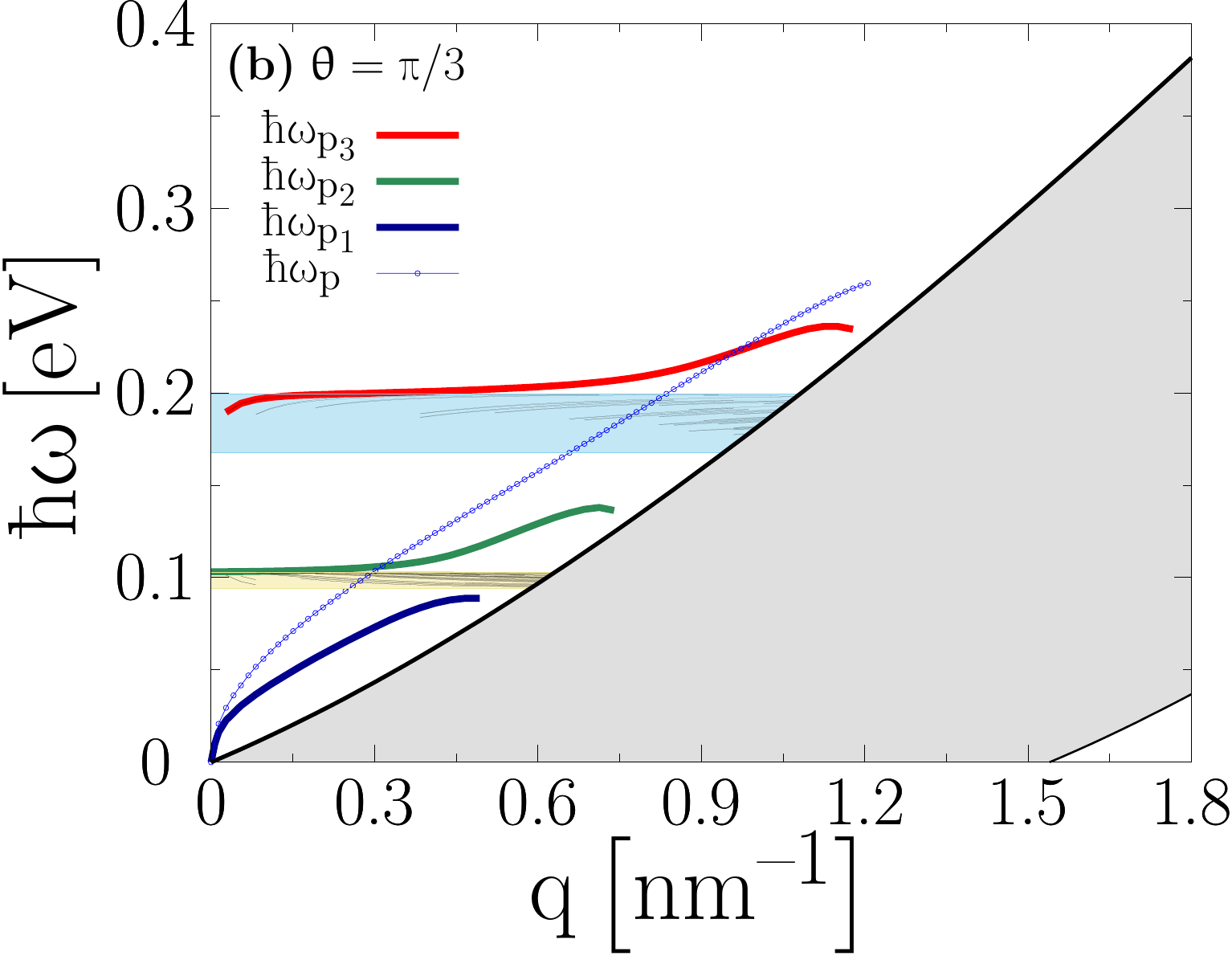}\\
	\hspace*{-0.2cm}
	\includegraphics[width=0.95\linewidth]{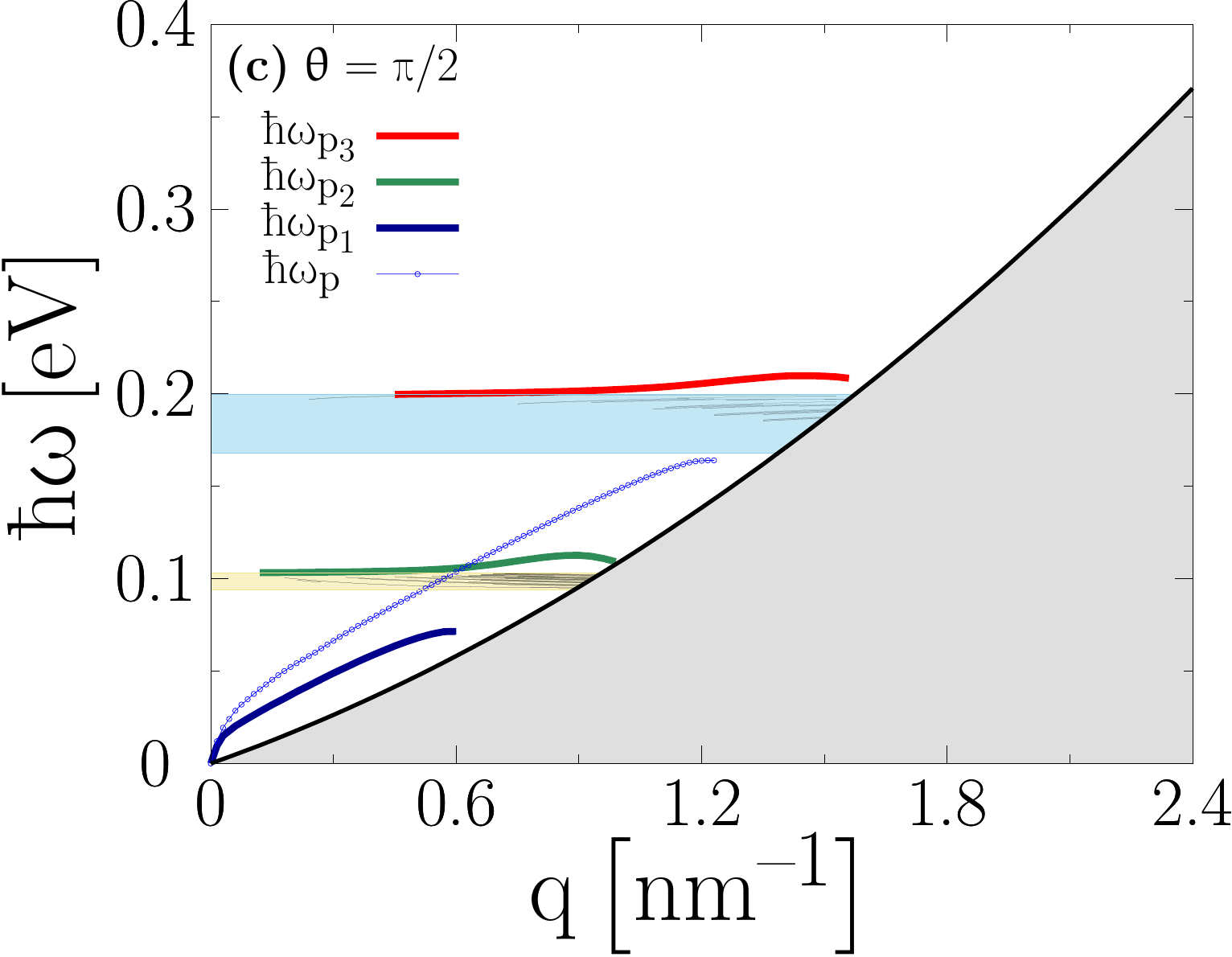}
	\caption{(Color online) Plasmon-phonon-polarities $\mathrm{\hbar \omega_{p_i}}$ (dark blue, green and red curves) and the corresponding simple plasmon $\mathrm{\hbar \omega_{p}}$ (blue curve) as a function of wavenumber $\mathrm{q}$ for electron doped system $\mathrm{n_e=1\times 10^{13} \, cm^{-2} \; (\mu=50.08 \, meV)}$ at room temperature along wavevector directions (a)~$\mathrm{\theta=0}$, (b)~$\mathrm{\theta=\pi / 3}$ and (c)~$\mathrm{\theta=\pi/ 2}$. The gray region is the intraband electron-hole continuum. The beige and sky-blue ribbons are the two Reststrahlen bands inherited from hBN with the energy intervals $\mathrm{[\hbar \omega_z^T , \hbar \omega_z^L ]}$ and $\mathrm{[\hbar \omega_x^T , \hbar \omega_x^L ]}$ in which there are numerous modes (monotonically decreasing/increasing thin black curves in the first/second band). Note the growth of the $\mathrm{q}$ scale and the decrease of the collective modes' energies by approaching to the zigzag direction.}
	\label{fig5}
\end{figure}
\begin{figure}
	\hspace*{-0.2cm}
	\includegraphics[width=0.95\linewidth]{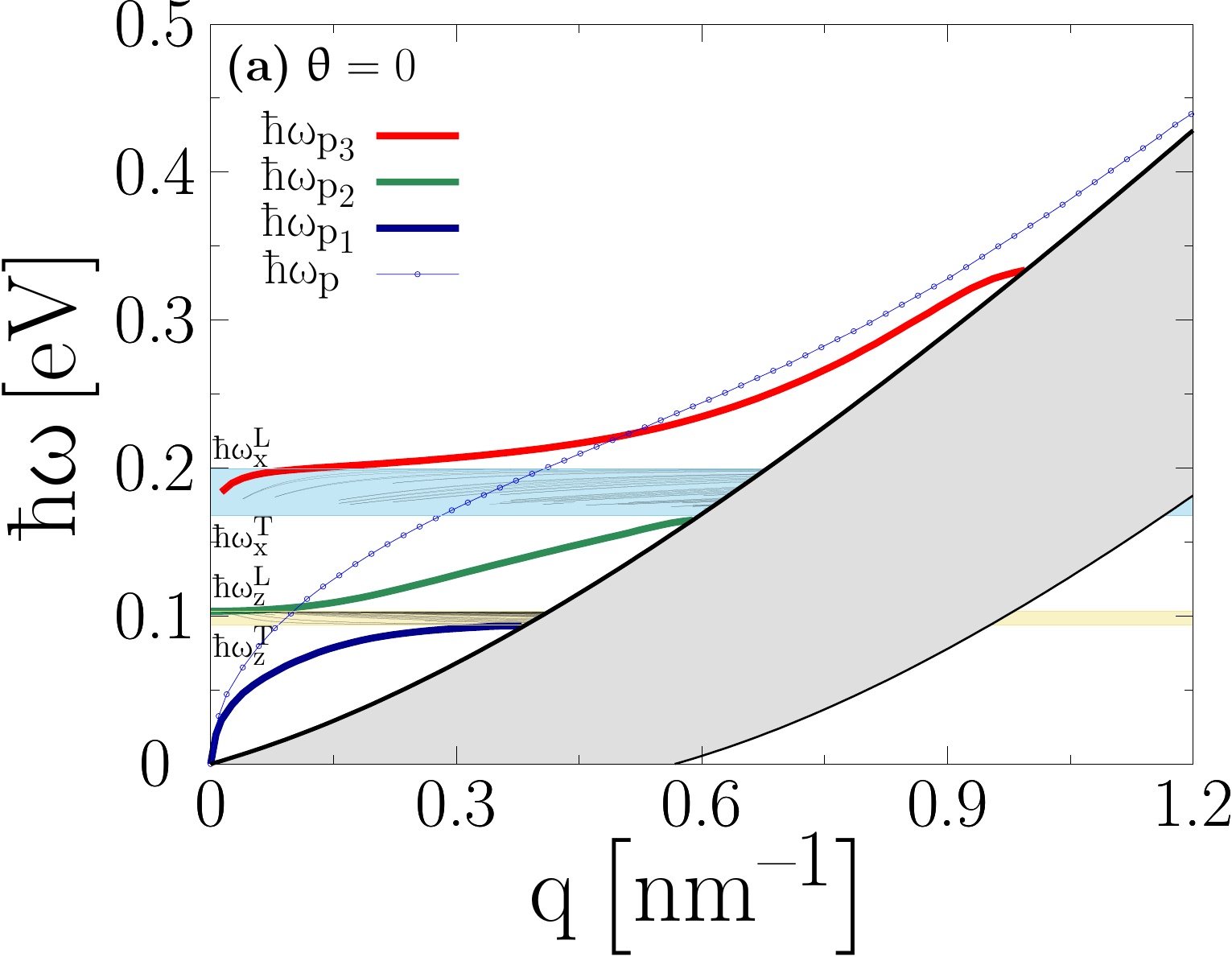}\\
	\hspace*{-0.2cm}
	\includegraphics[width=0.95\linewidth]{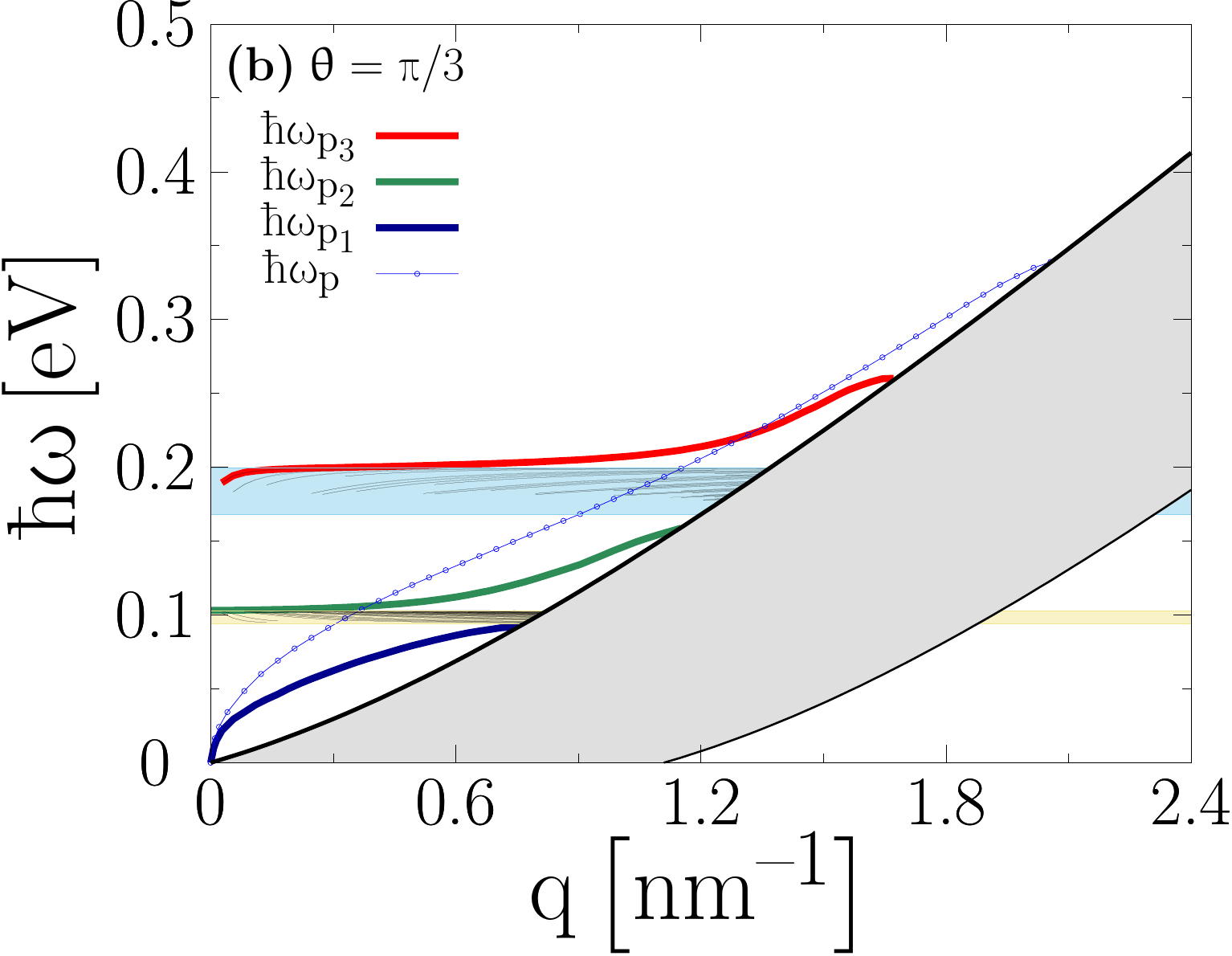}\\
	\hspace*{-0.2cm}
	\includegraphics[width=0.95\linewidth]{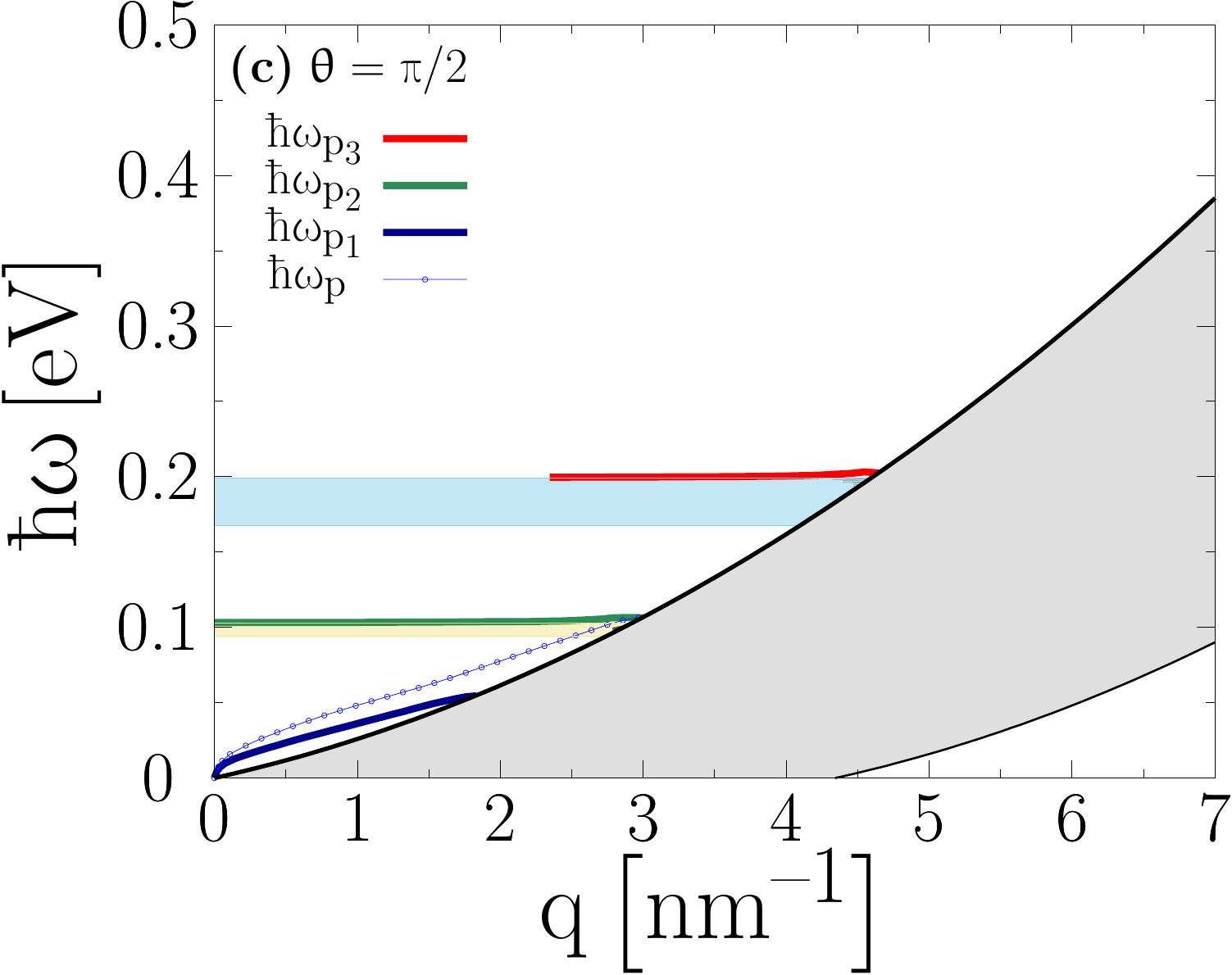}
	\caption{(Color online) Plasmon-phonon-polarities $\mathrm{\hbar \omega_{p_i}}$ (dark blue, green and red curves) and the corresponding simple plasmon $\mathrm{\hbar \omega_{p}}$ (blue curve) as a function of wavenumber $\mathrm{q}$ for hole doped system $\mathrm{n_h=1\times 10^{13} \, cm^{-2} \; (\mu=-22.87 \, meV)}$ at $\mathrm{T=50\, K}$ along wavevector directions (a) $\mathrm{\theta=0}$, (b)~$\mathrm{\theta=\pi / 3}$ and (c) $\mathrm{\theta=\pi/2}$. For details see Fig. \ref{fig5}. The $\mathrm{q}$ scale's growth of the dispersions and their energy decrease by approaching to the zigzag direction are more drastic here than in Fig. \ref{fig5}.}
	\label{fig6}
\end{figure}

Having calculated the noninteracting charge-charge linear response function at finite temperature, we could calculate the many-body dielectric function and accordingly the plasmon modes. In order to assess the approach and the model Hamiltonian, we first compare the plasmon mode of a free-standing phosphorene with that of calculated in \cite{TorbatianPRB} based on the density functional theory (DFT)-RPA approach at zero temperature for $n_e=3.5\times 10^{13}$ cm$^{-2}$. The calculated plasmon mode of the system along the armchair and zigzag directions are depicted in Fig. \ref{fig4} \cite{note,Torbatian}. 

Now, we would like to consider our studied system depicted in Fig. \ref{fig2}. We study the behavior of the simple plasmon modes $\hbar \omega_p$ in our simple model and that of the SPPP modes $\hbar \omega_{p_i}$ (with $i=1,2,3$) in our coupled model in two physical conditions: (1) an electron doped system at room temperature $T = 300 \,$K with the electron density $n_e=1 \times 10^{13}$ cm$^{-2}$ (corresponding to the chemical potential $\mu = 50.08$ meV), and (2) a hole doped system at $T = 50 \,$K with the hole density $n_h=1 \times 10^{13}$ cm$^{-2}$ (corresponding to the chemical potential $\mu = -22.87$~meV).

The dispersions of those plasmon modes (all in units of eV) as a function of the wavenumber $q$ (in units of nm$^{-1}$) along three wavevector directions; namely (a) the armchair direction for $\theta = 0$, (b) a finite angle direction for $\theta = \pi /3 $ and (c) the zigzag direction for $\theta=\pi/2$ for the two physical conditions are presented in Figs. \ref{fig5} and \ref{fig6}, respectively. In these figures the simple plasmon modes $\hbar \omega_p$ are shown by blue curves while the three SPPP modes $\hbar \omega_{p_i}$ are shown respectively by dark blue, green and red curves. As said before, the electron-hole continuum with boundaries $\hbar \omega_\pm^\tau (\textbf{q})  = \varepsilon^\tau_{\textbf{q} \pm \textbf{k}_F} - \varepsilon^\tau_{\textbf{k}_F}$ is illustrated by a gray region. The beige and sky-blue horizontal ribbons are the two Reststrahlen bands inherited from hBN with the energy intervals $[\hbar \omega_z^T , \hbar \omega_z^L ]$ and $[\hbar \omega_x^T , \hbar \omega_x^L ]$, respectively.

The plasmon modes exhibit a strong anisotropy. Furthermore, the significant feature visible in the results (in addition to the anisotropy) is the complete reshaping of the typical square root dispersion of the bare plasmon mode to a three-branched dispersion (SPPPs) in the vicinity of the frequency of the hBN's optical phonon. The amount of phonon-like or plasmon-like content can be qualitatively inferred by surveying the spectrum. The lowest branch is clearly plasmon-like which behaves like a simple and uncoupled phosphorene plasmon mode. In the long-wavelength limit, two other modes show phonon-like behavior.
\begin{figure*}
    \hspace*{-0.23cm}
	\includegraphics[width=0.497\linewidth]{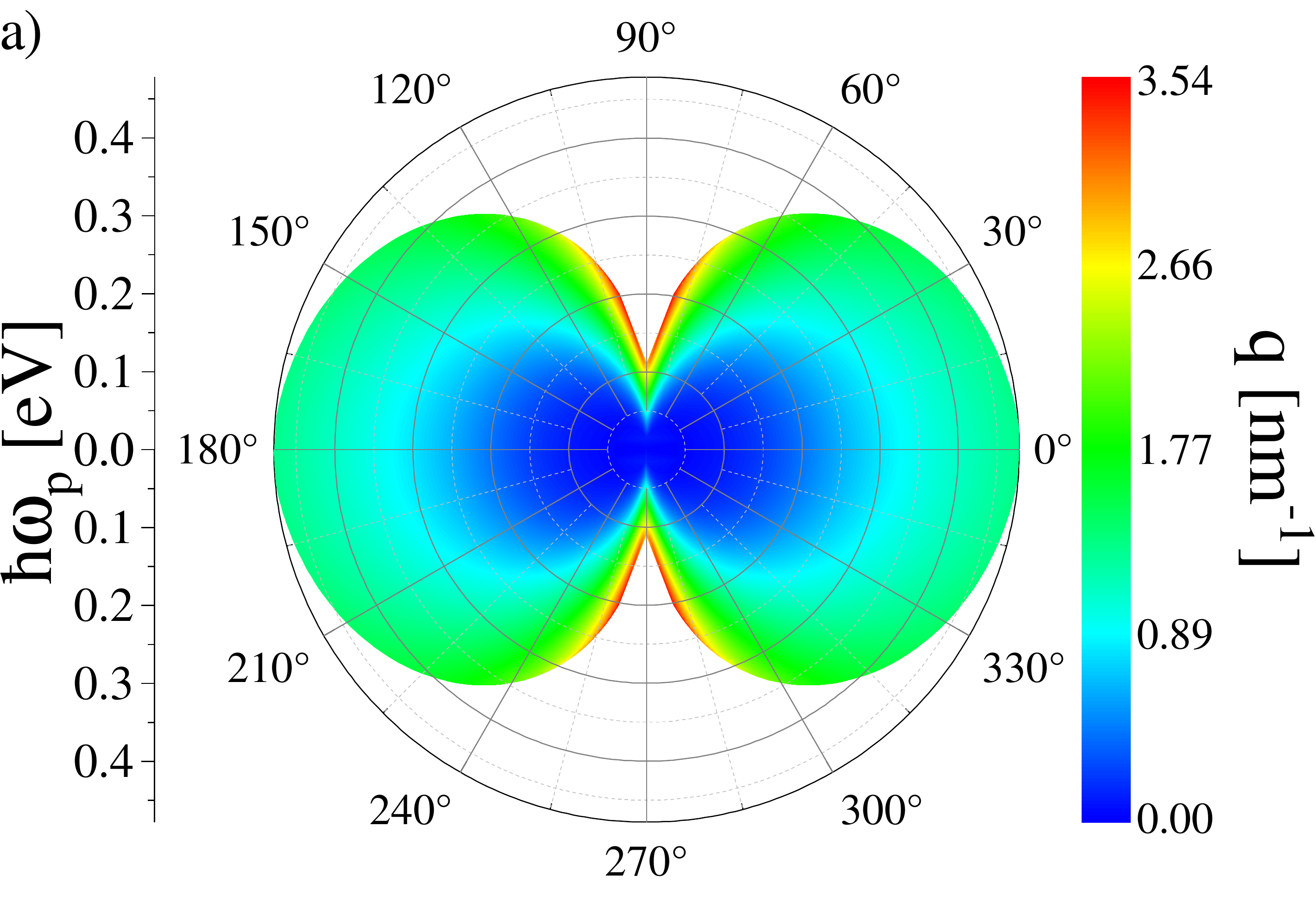}
	\includegraphics[width=0.497\linewidth]{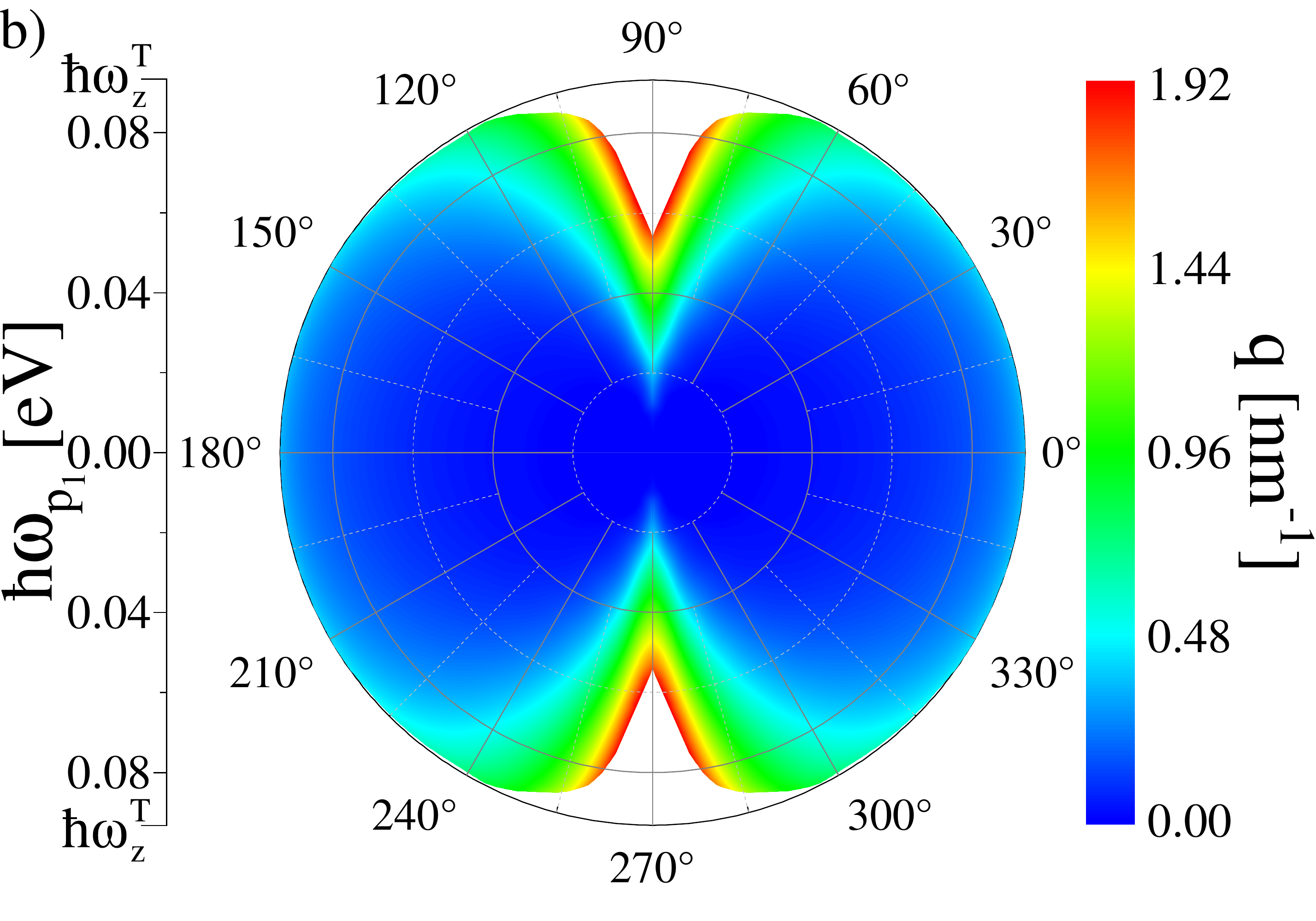}\\
	\hspace*{-0.34cm}
	\includegraphics[width=0.497\linewidth]{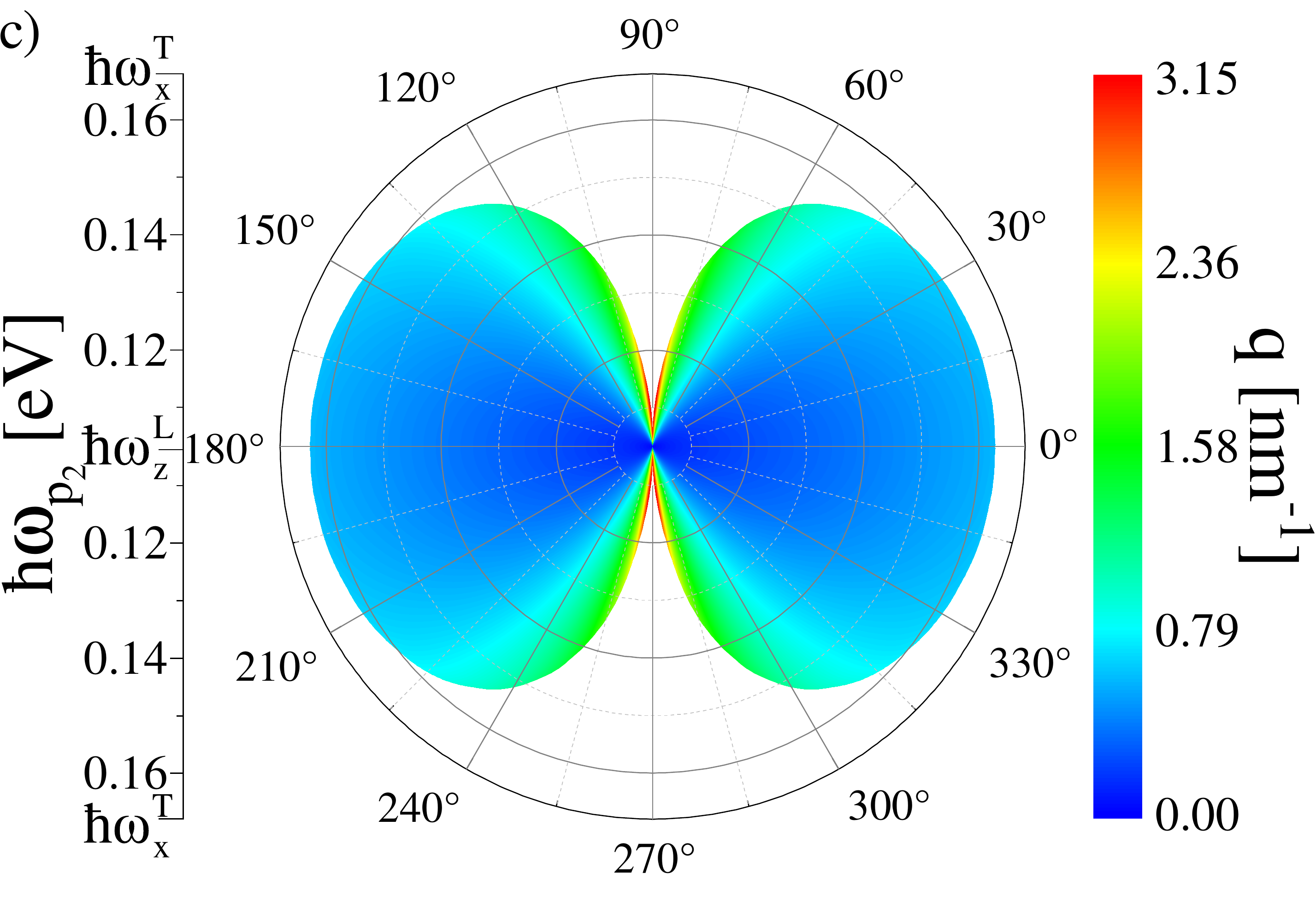}
	\includegraphics[width=0.497\linewidth]{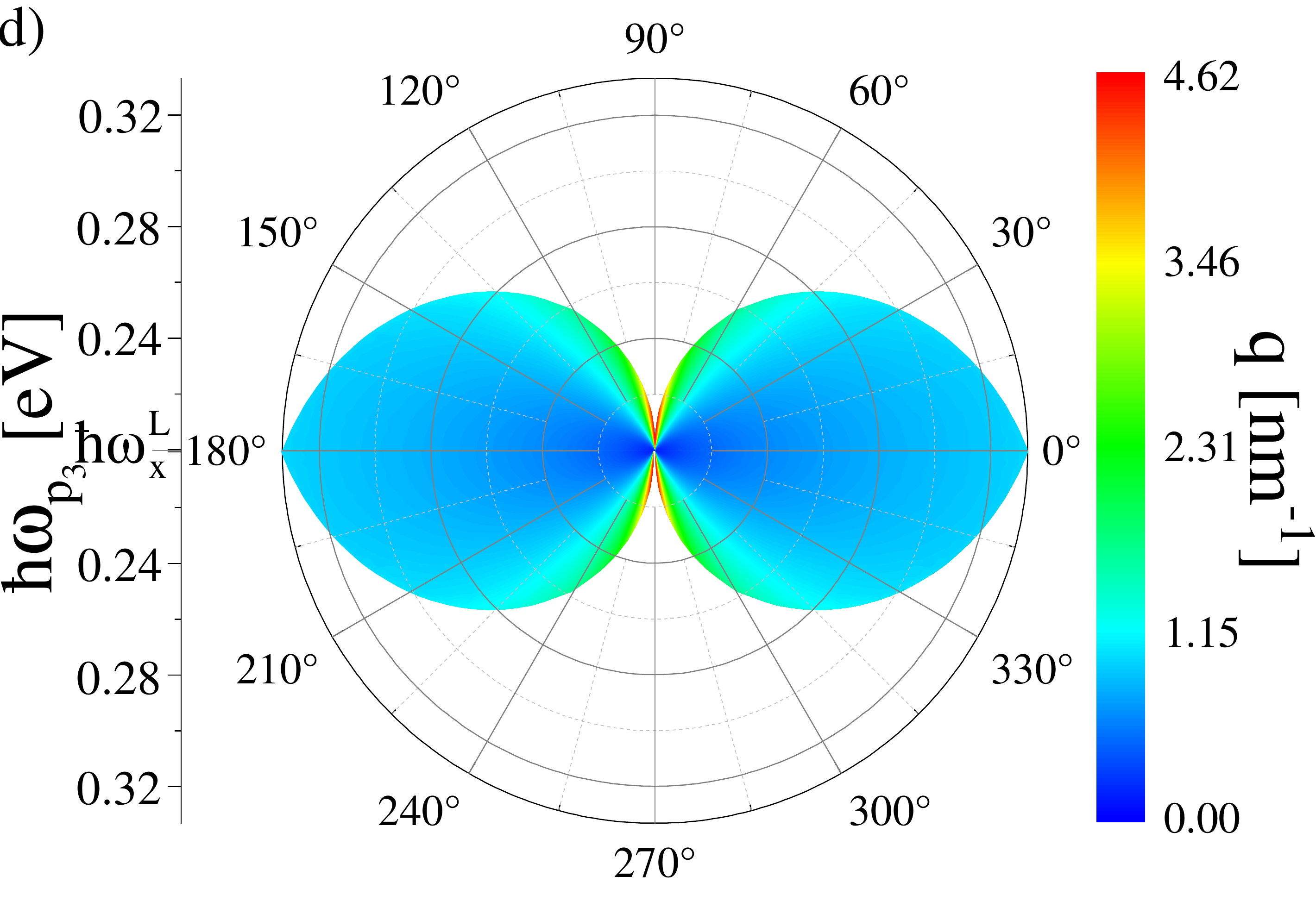}
	\caption{(Color online) Polar contour plots of (a) simple plasmon $\mathrm{\hbar\omega_p}$ and (b-d) plasmon-phonon-polarities $\hbar\omega_{p_i}$ for the physical condition described in Fig. \ref{fig6}. Energies are put at radius and  wavenumbers are illustrated by color due to high anisotropy between the armchair and the zigzag directions. In (b) $\mathrm{\hbar\omega_{p_1}}$ is limited to $\mathrm{\hbar\omega_z^T}$ from the upper bound after which is the first Reststrahlen band $\mathrm{[\hbar\omega_z^T,\hbar\omega_z^L]}$. In (c) $\mathrm{\hbar\omega_{p_2}}$ is confined to $\mathrm{\hbar\omega_z^L}$ and $\mathrm{\hbar\omega_x^T}$ (respectively) from the lower and the upper bound. $\mathrm{\hbar\omega_{p_3}}$ has a tail in the second Reststrahlen band $\mathrm{[\hbar\omega_x^T,\hbar\omega_x^L]}$ but in (d) we ignore the beginning tail and just study and depict the upper part of it which is higher than $\mathrm{\hbar\omega_x^L}$.}
	\label{fig7}
\end{figure*}
\begin{figure*}
	\centering
	\includegraphics[width=3.4in]{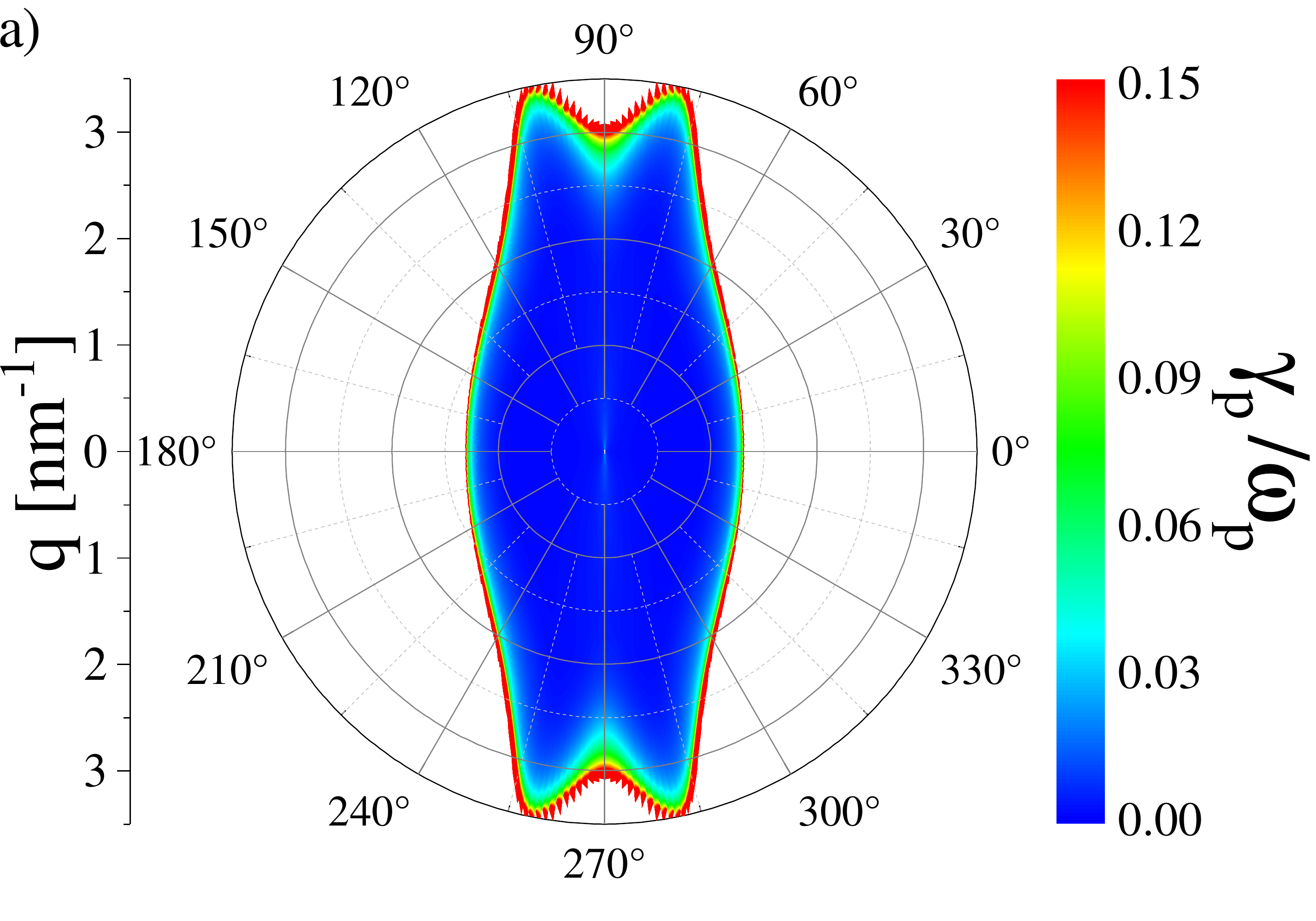}
	\includegraphics[width=3.4in]{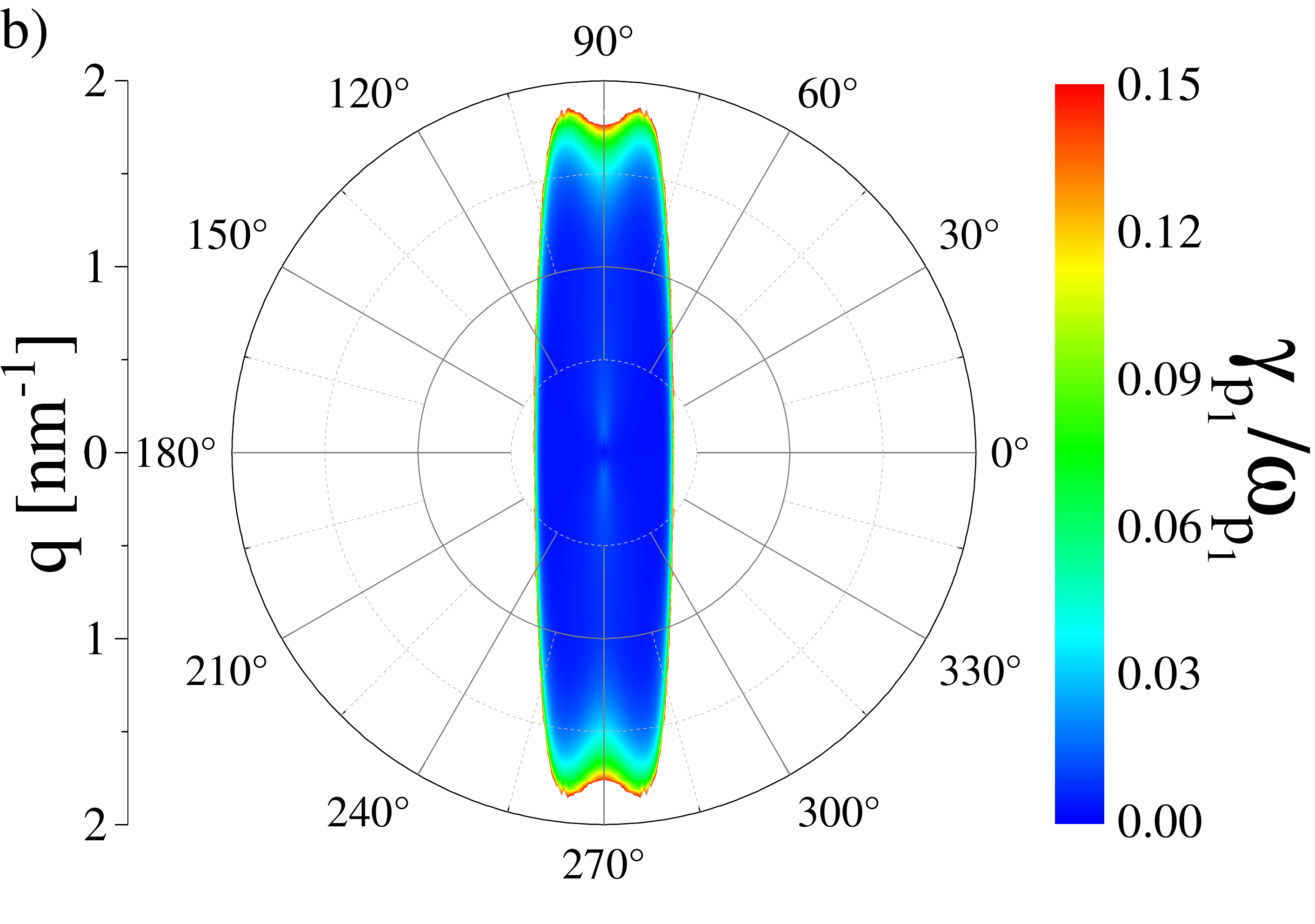}
	\includegraphics[width=3.4in]{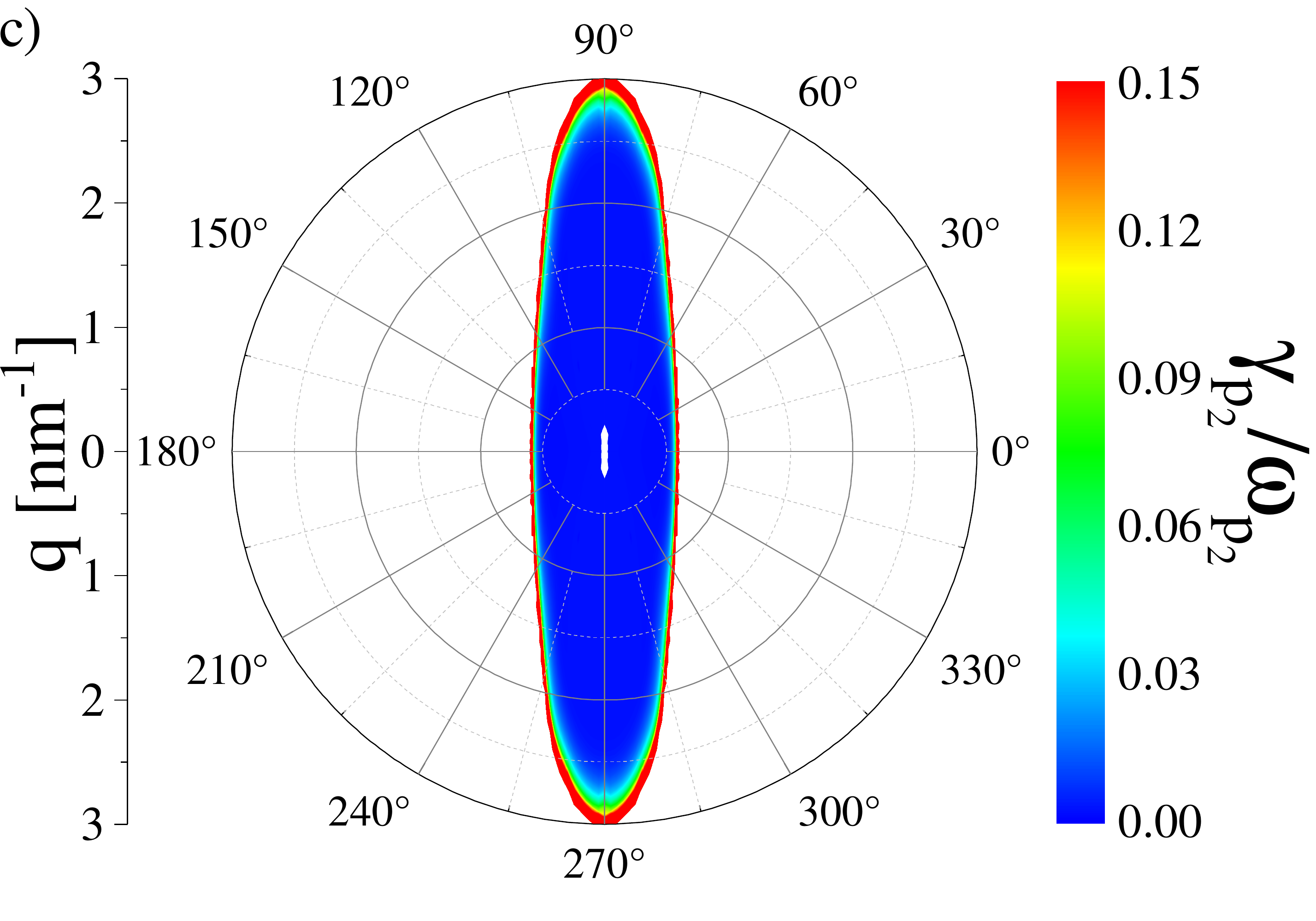}
	\includegraphics[width=3.4in]{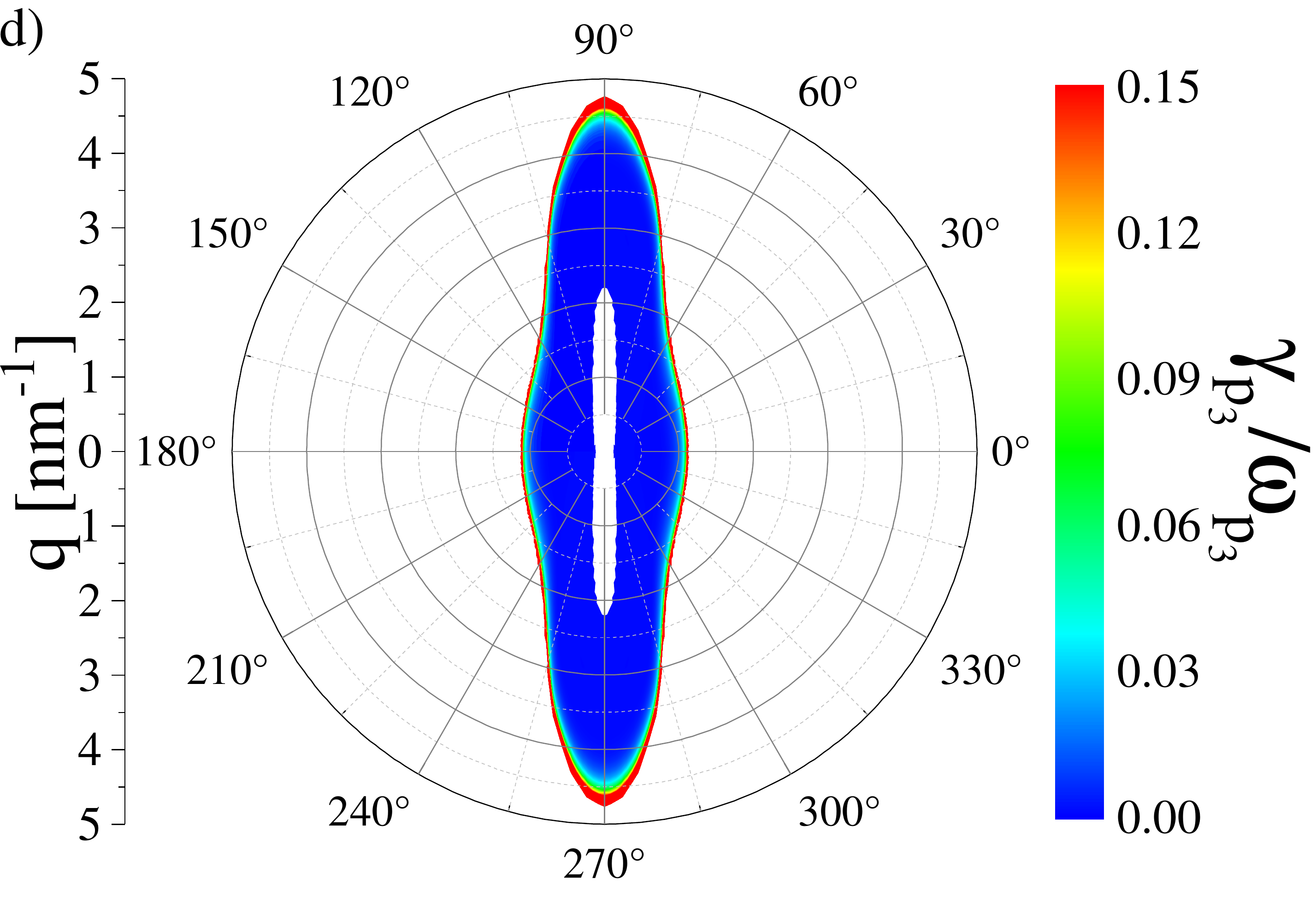}
	\caption{(Color online) Polar contour plots of (a) $\mathrm{\gamma_{p} / \omega_p}$, the rescaled damping parameter of the simple plasmons in units of the plasmon angular frequency and (b-d) $\mathrm{\gamma_{p_i} / \omega_{p_i}}$, the rescaled damping parameter of the plasmon-phonon-polarities in units of the SPPP angular frequencies. The physical conditions are identical to those of Fig. \ref{fig6} and Fig. \ref{fig7}.}
	\label{fig8}
\end{figure*}

The simple plasmon mode $\hbar \omega_p (\mathbf{q})$ depicted by blue curves, behaves differently in terms of the wavevector angle $\theta$ owing to their mass anisotropy where the smaller mass along the $k_x$ direction leads to higher resonance frequency. Juxtaposition of the two Figs. \ref{fig5} and \ref{fig6} in the electron and hole doped cases, exhibits much stronger anisotropy of the plasmon modes of the hole doped system in comparison with those of the electron doped system. Moreover, the Landau damping occurs due to the intraband processes when plasmon enters intraband electron-hole continuum (gray region). By comparison of Figs. \ref{fig5}(a) and \ref{fig6}(a) along $\theta=0$ with Figs. \ref{fig5}(b) and \ref{fig6}(b) along $\theta=\pi / 3$, one observes a smooth energy decreasing of the plasmon mode by increasing of the wavevector angle. Overall comparison of the two Figs. \ref{fig5} and \ref{fig6} reconciles the fact that this decreasing of the plasmon energy is more intense in the hole doped case.

We also calculate the coupling of plasmon with optical phonon modes of the polar dielectric environment. Remarkably, this coupling is strong at all charge densities. In ordinary 2D systems, the plasmon-phonon mode coupling is only significant at densities satisfying the resonance condition where the value of the plasmon mode is close to that of the optical phonon. However, the collective electronic excitations not only exist for all wavevectors in encapsulated phosphorene, but also they exhibit a strong in-plane anisotropy. As seen in Figs. \ref{fig5} and \ref{fig6}, the SPPP mode decomposes into three branches $\hbar \omega_{p_i}$ which are illustrated in the figures by dark blue curves for the first gapless branch ($i=1$) with the upper limit of the first Reststrahlen band $[\hbar \omega_z^T , \hbar \omega_z^L]$, by green curves for the second branch ($i=2$) between the first Reststrahlen band and the second one $[\hbar \omega_x^T , \hbar \omega_x^L ]$  and ultimately by red curves for the third branch ($i=3$) launched at the middle of this second Reststrahlen band. These three branches of the SPPP modes are dependent on the momentum direction and their anisotropy is again much stronger in the hole-doped case in the same way of the plasmon mode, as it is inferred from a comparison between the triple panels of the figures due to triple selected wavevector directions. It can be inferred from Figs. \ref{fig5}(c) and \ref{fig6}(c), the two upper SPPPs are so phonon-like along the zigzag direction $\theta=\pi/2$.

In addition to these three SPPP modes in Figs. \ref{fig5} and \ref{fig6}, there are numerous monotonically decreasing (increasing) modes inside the first (second) Reststrahlen band illustrated by thin black curves.

Aiming to illustrate the highly anisotropic nature of the plasmon mode $\hbar \omega_p (\textbf{q})$ and the SPPP modes $\hbar \omega_{p_i} (\textbf{q})$ of the encapsulated phosphorene discussed above, their contour plots in the hole doped case are included in Figs. \ref{fig7}(a) and \ref{fig7}(b-d), respectively. All the physical conditions considered in Fig. \ref{fig7} is identical to those of Fig. \ref{fig6}. To avoid unpleasant and vague tight long elliptical contour plots appeared in Fig. \ref{fig8} owing to a high anisotropy between the armchair and the zigzag directions, the energies of the plasmon modes (in units of eV) are put at radius and the wavenumbers (in units of nm$^{-1}$) are illustrated by color. Considering wavenumbers in colors, it can be inferred from the figure that the anisotropy accelerates after an angular turning point which is about $60^\circ$ up to $70^\circ$. Analysis of the contour plot shapes which are pertinent to the radial energy dispersions, disclose the previously described facts that the damping of the collective modes are also directional-dependent. This angular turning point of the collective mode's damping depends on the mode and the value of the Fermi energy.
 
The plasmon energy decreases by increasing the wavevector angle $\theta$ up to around 75$^\circ$ and then experiences an abruptly intensified decreasing up to the zigzag direction 90$^\circ$. Therefore, it is wise to say that the plasmon energy along the $k_y$ direction is decreased abruptly at mid-infrared frequencies which is interesting from both fundamental and technological points of view \cite{Badioli}. Apparently, the SPPP modes of holes experience a much stronger decreasing energy in the zigzag direction in comparison with those of electrons due to the stronger anisotropy of the band effective masses and their dispersions.

On the other hand, it is helpful here to calculate the damping parameter $\gamma (\textbf{q}, \omega_p(\textbf{q}))$ of any plasmon mode $\hbar\omega_p(\textbf{q})$ at each wavevector $\textbf{q}$ by using Eq. (\ref{damping}). As mentioned before, its rescaled manifestation in (units of the collective mode's angular frequency) $\gamma (\textbf{q}, \omega_p(\textbf{q})) / \omega_p(\textbf{q})$ or simply $\gamma_p / \omega_p$ is a robust signifier of the damping intensity of the collective mode under consideration at that wavevector. This rescaled damping parameter is calculated in both cases of the simple and the coupled model and the results are illustrated in Fig. \ref{fig8} as polar contour graphs with $q$ (in units of nm$^{-1}$) in radius and with the damping parameter in colors. Fig. \ref{fig8}(a) depicts $\gamma_p / \omega_p $ for the simple plasmon mode in the simple model while the Figs. \ref{fig8}(b-d) represent $\gamma_{p_i} / \omega_{p_i}$ for the SPPP modes in the coupled model. This shows that while all of these modes are highly anisotropic and are elongated in $q$ by increasing the wavevector angle from the armchair direction $\theta=0^{\circ}$ up to the zigzag direction $\theta=90^{\circ}$, they are much smaller than unity, and therefore they are not damped except in the vicinity of the upper boundary of the intraband electron-hole continuum. But since the $\Im m \, \varepsilon(q,\omega)$ at the position of the plasmon modes are finite, it turns out that those collective modes are damped and the damping parameter $\gamma$ as a function of the momentum would be small. It is worth noting that the simple mode in Fig. \ref{fig8}(a) and the first SPPP mode in Fig. \ref{fig8}(b) are both gapless and possess the most wavevector-elongated modes not at the zigzag direction but at angle about $\theta=78^{\circ}$ and $\theta=82^{\circ}$, respectively.
%
\subsection{RPA and Semiclassical Model (SC) Comparison}
\label{sec3a}
%
\begin{figure}
  \centering
	\includegraphics[width=\linewidth]{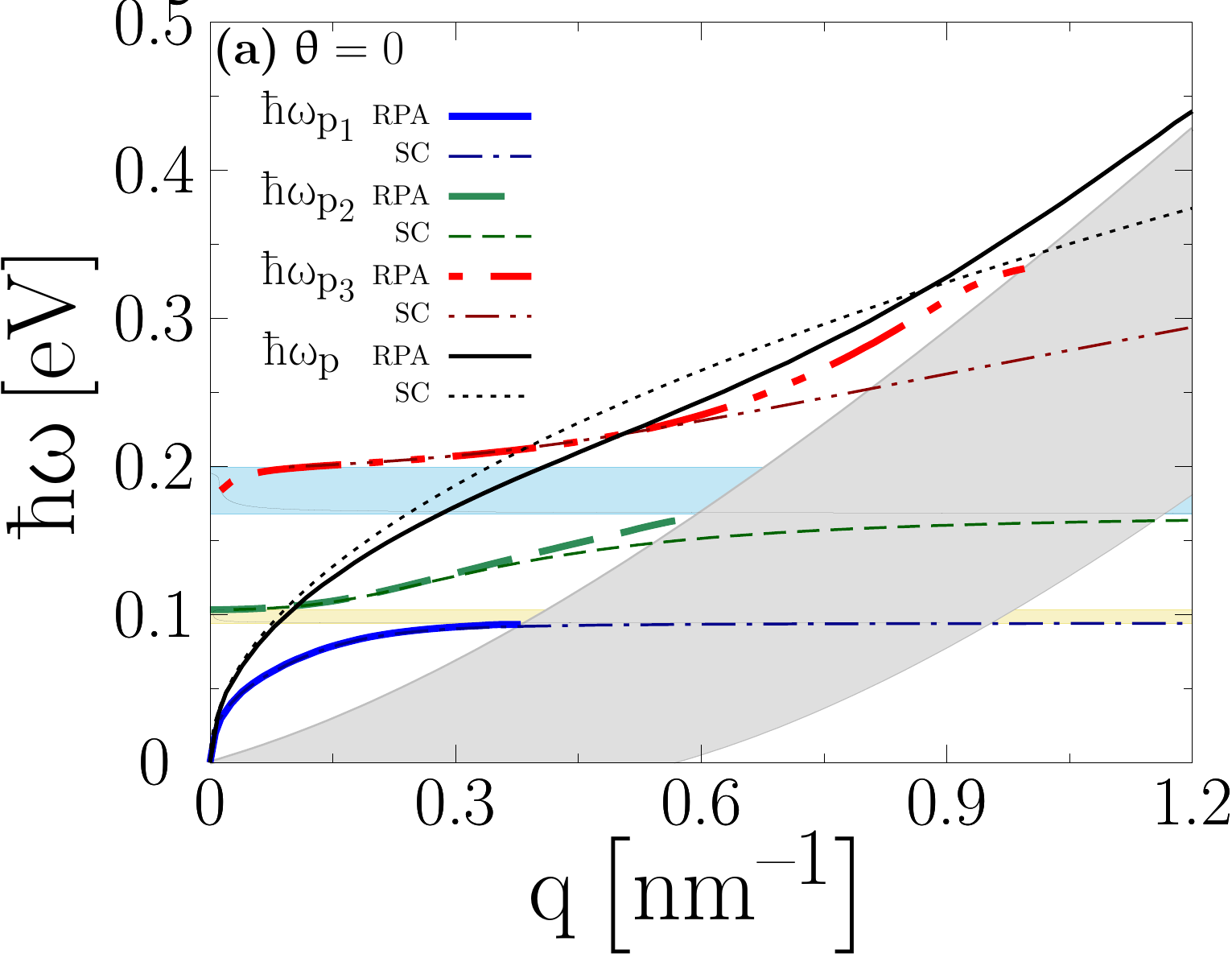}
	\hspace*{-0.44cm}
	\includegraphics[width=\linewidth]{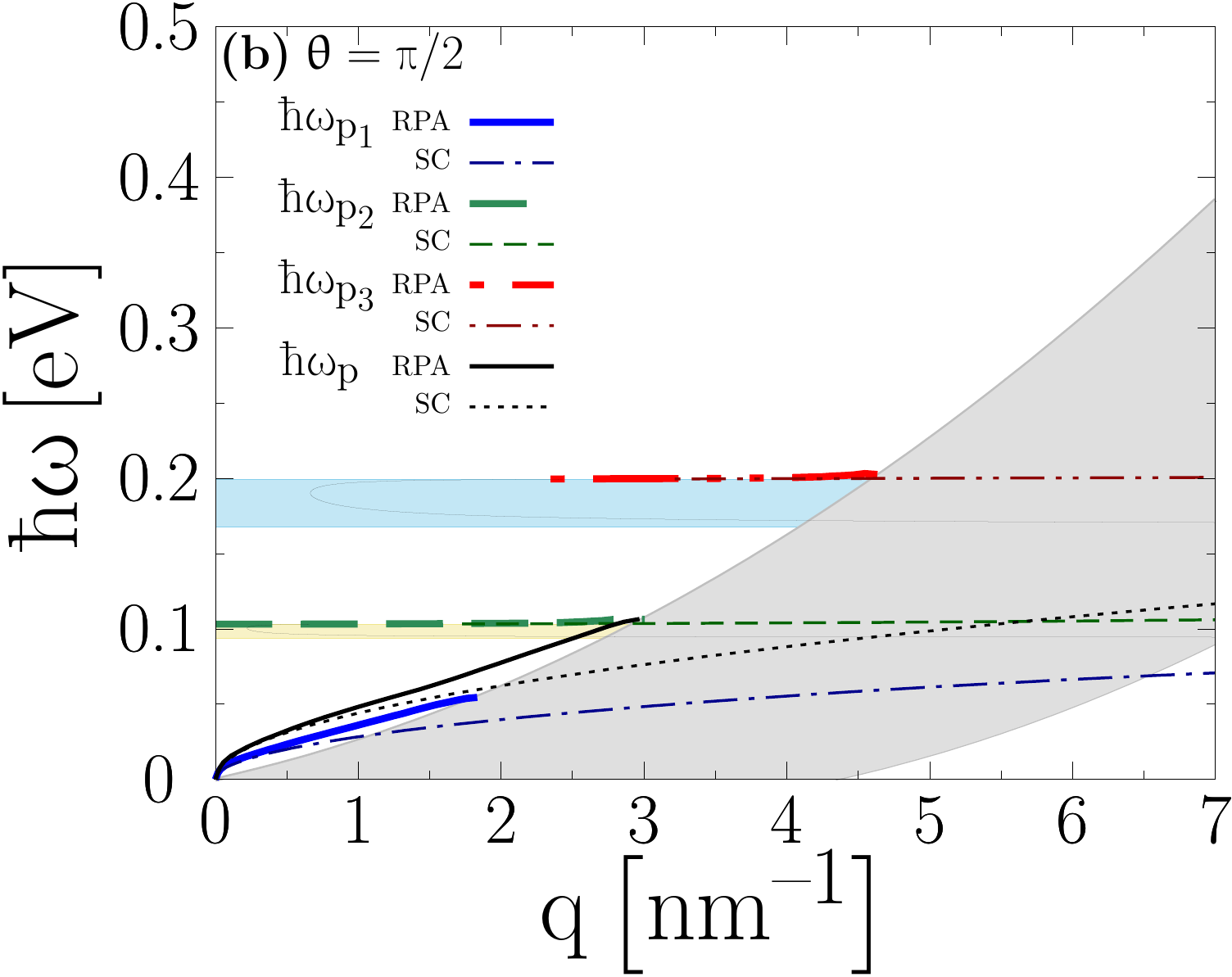}
	\caption{(Color online) The semiclassical (SC) dispersions of the surface plasmon-phonon-polaritons (thin dashed curves in blue, green and red color) and the corresponding simple plasmon (thin dotted black curve) in comparison with the RPA counterparts (thick curves) along wavevector directions (a) $\mathrm{\theta=0}$ and (b)~$\mathrm{\theta=\pi/2}$ for the system indicated at Fig. \ref{fig6} in which the RPA results was initially presented.}
	\label{fig9}
\end{figure}
\begin{figure*}
  \centering
	\includegraphics[width=0.495\linewidth]{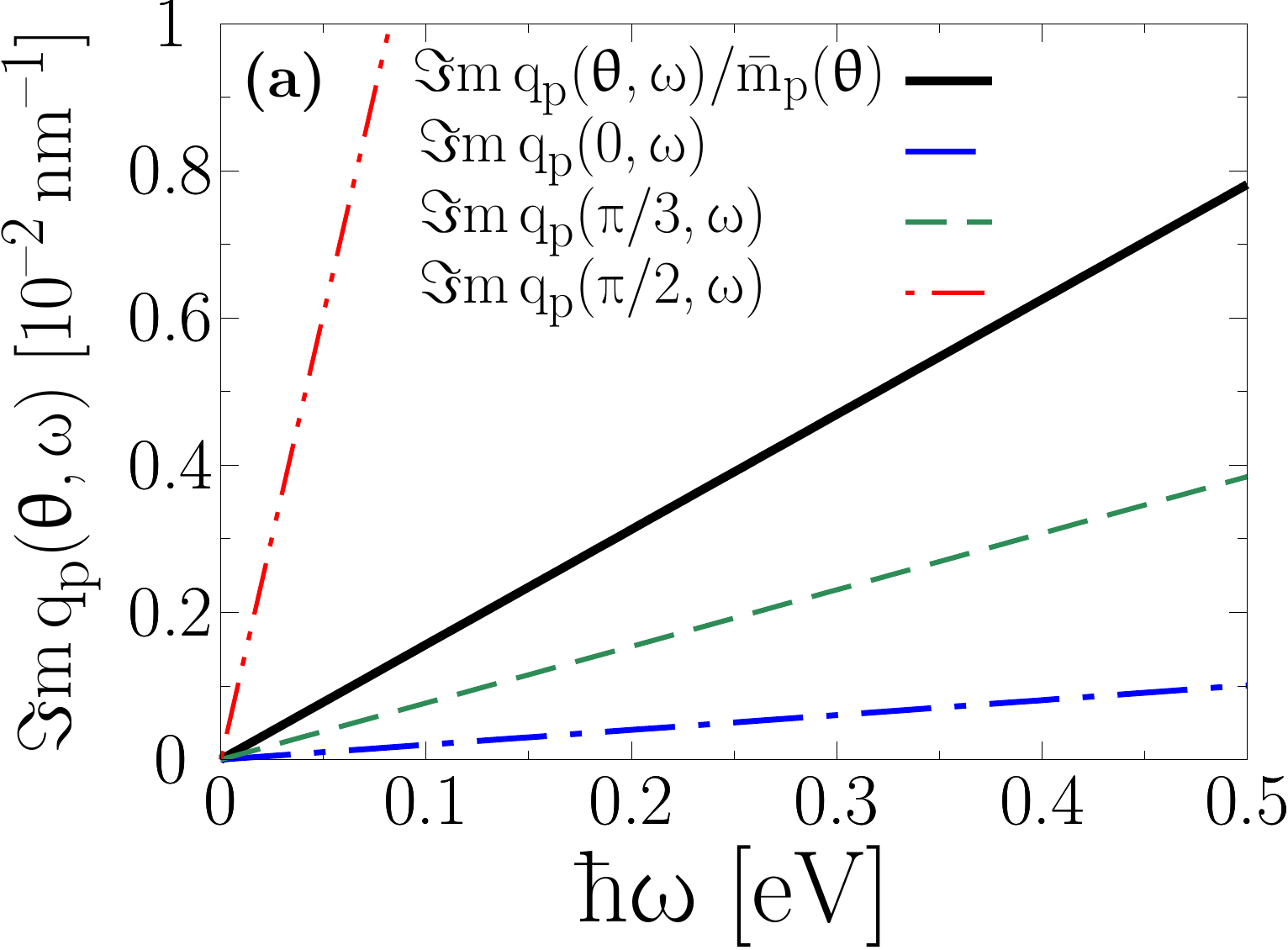}
	\includegraphics[width=0.495\linewidth]{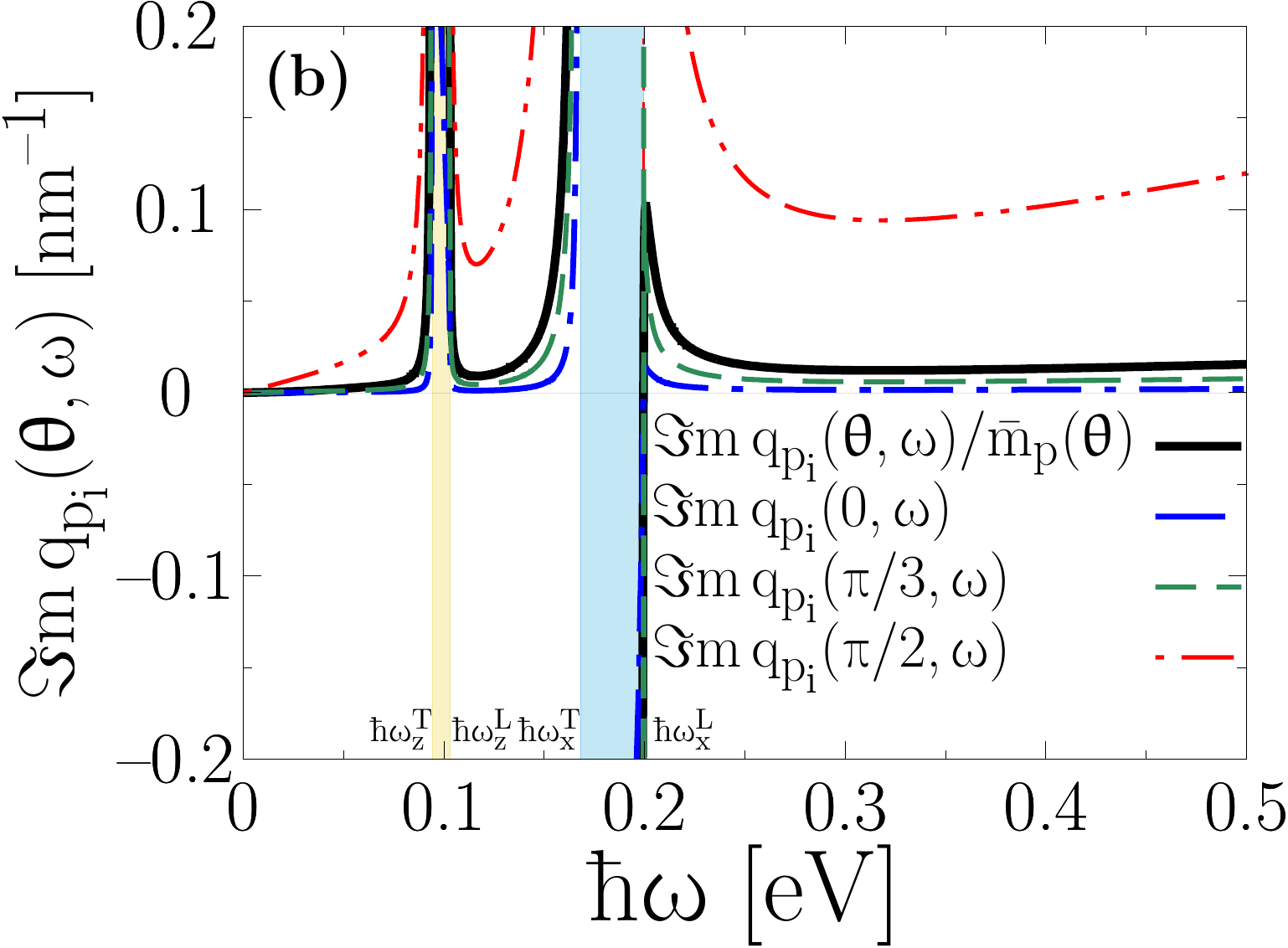}
	\includegraphics[width=0.495\linewidth]{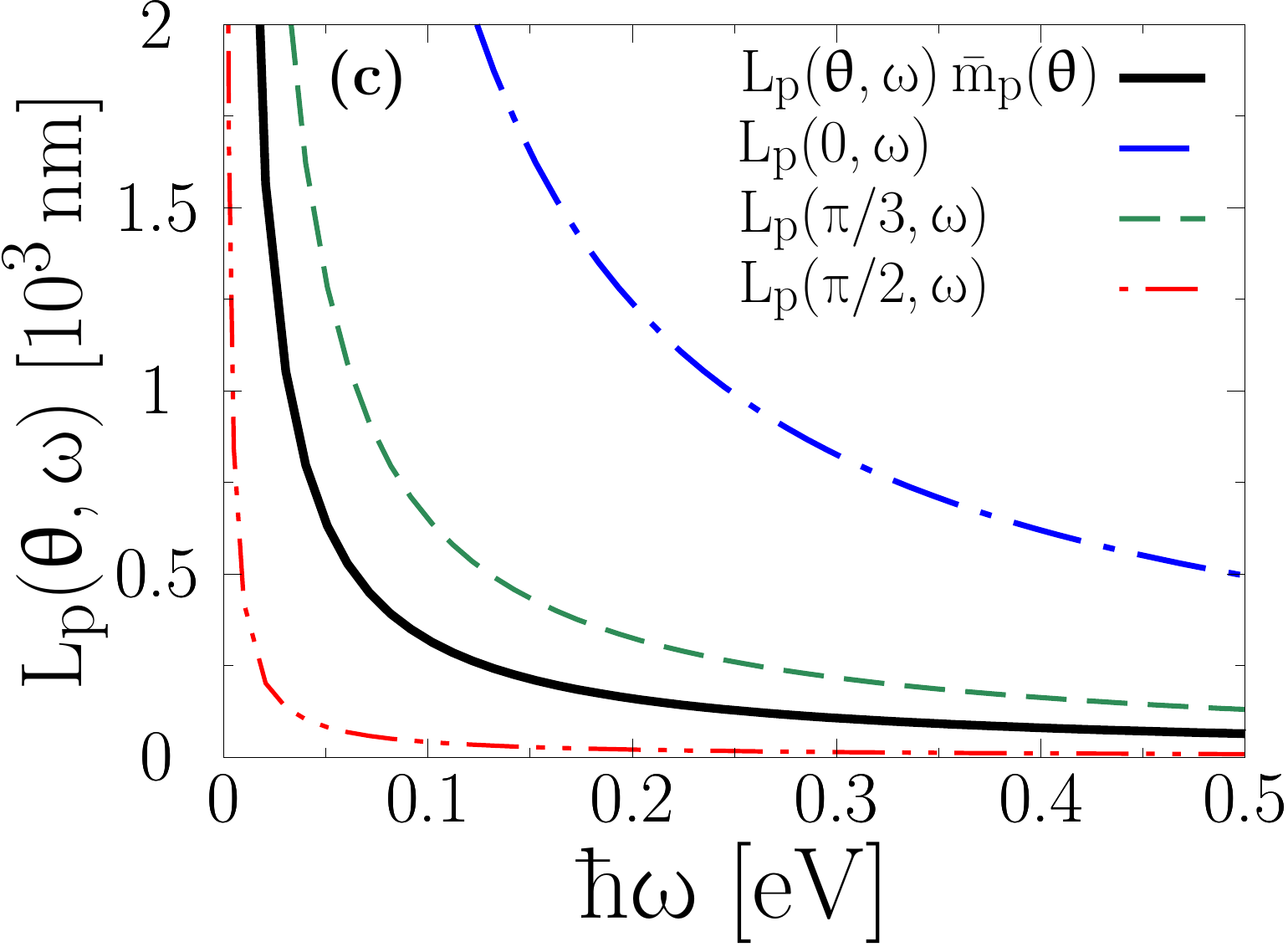}
	\includegraphics[width=0.495\linewidth]{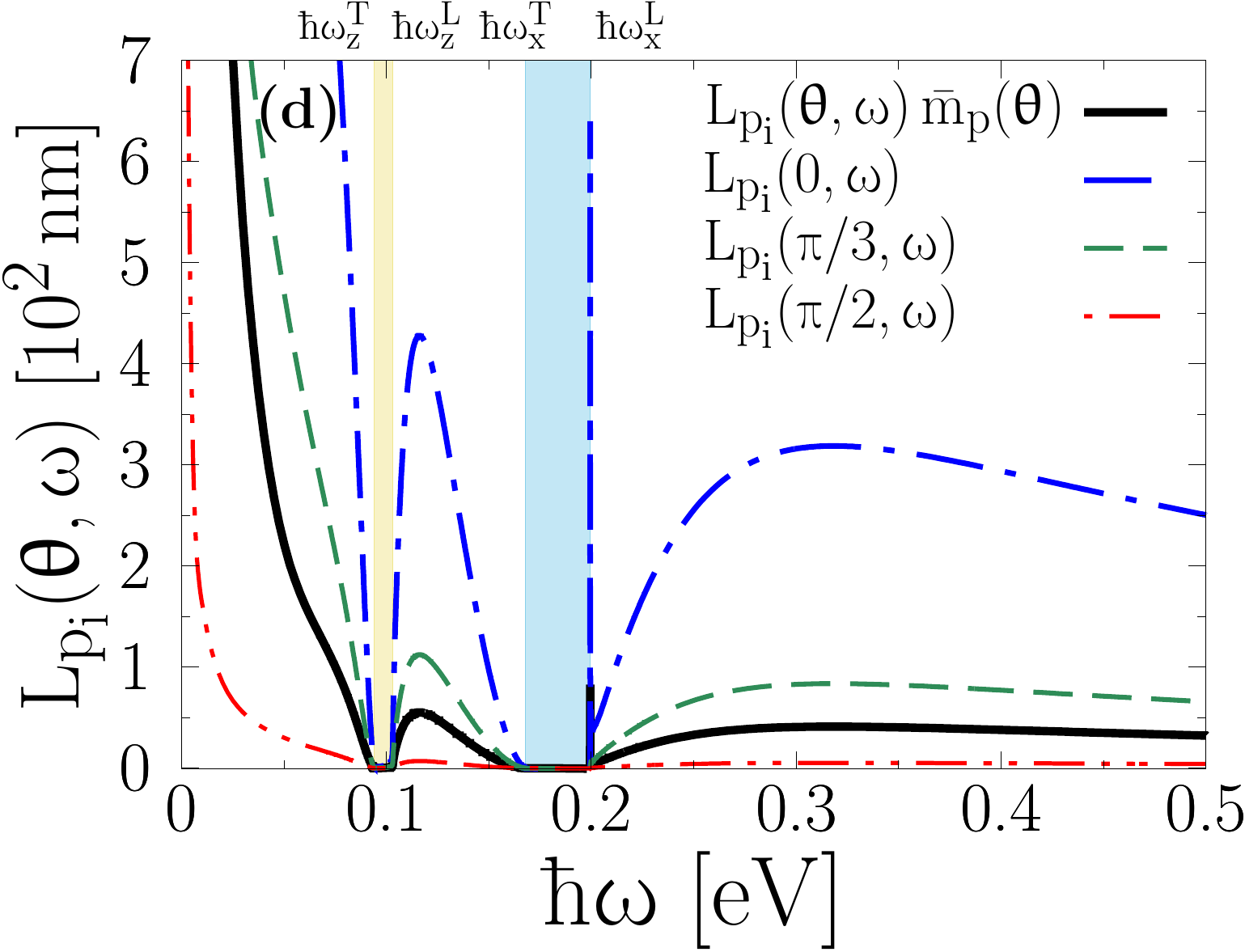}
	\includegraphics[width=0.495\linewidth]{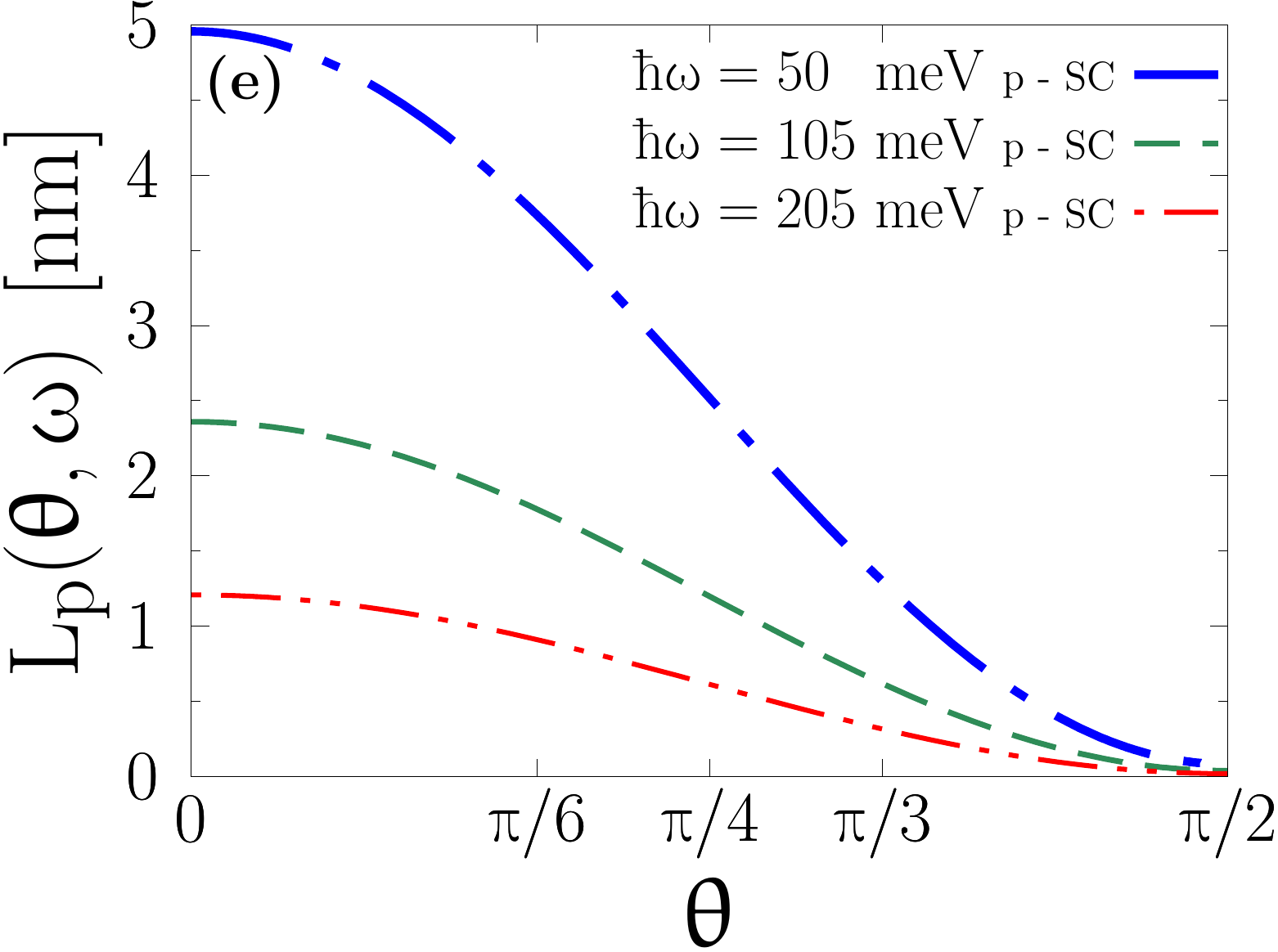}
	\includegraphics[width=0.495\linewidth]{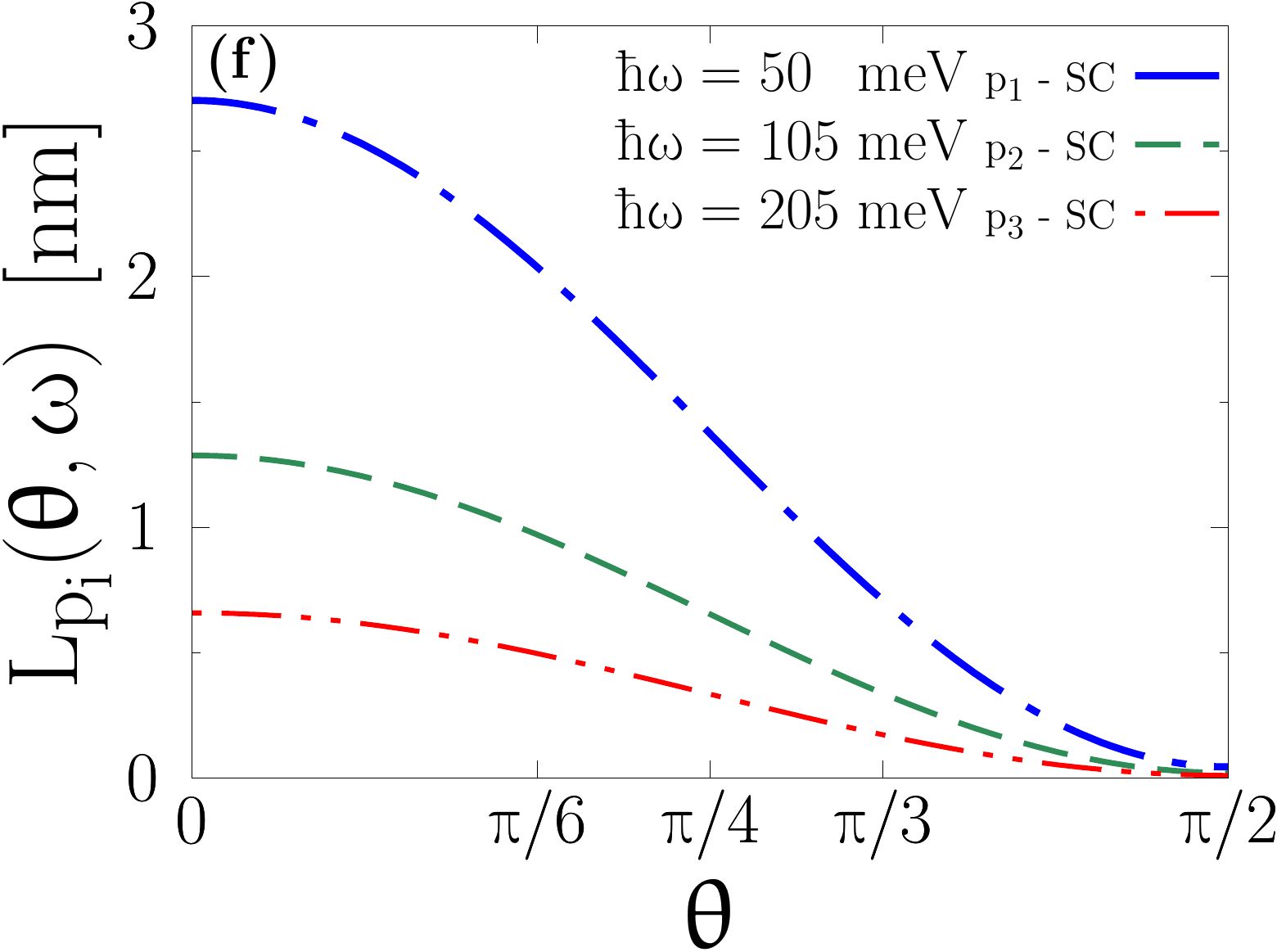}
	\caption{(Color online) The imaginary part of the wavevector in the semiclassical model for (a) the simple plasmon and (b) the surface plasmon-phonon-polaritons as a function of frequency along wavevector angles $\mathrm{\theta=0,\pi/3,\pi/2}$ for the system indicated in Fig. \ref{fig6}. Their respective propagation lengths $\mathrm{L_{p/p_i}(\omega,\theta)}$ are illustrated in (c) and (d).  The thick black curves in panels (a-d) are scaled curves relative to the dimensionless directional mass $\mathrm{\bar{m}_p(\theta)=m_p(\theta)/m_d}$. The angular dependency of the propagation lengths for three selected frequencies $\mathrm{\hbar \omega =50, 105, 205 \; meV}$ is depicted in (e) for simple plasmon and in (f) for surface plasmon-phonon-polaritons.}
	\label{fig10}
\end{figure*}
\begin{figure*}
  \centering
	\includegraphics[width=0.492\linewidth]{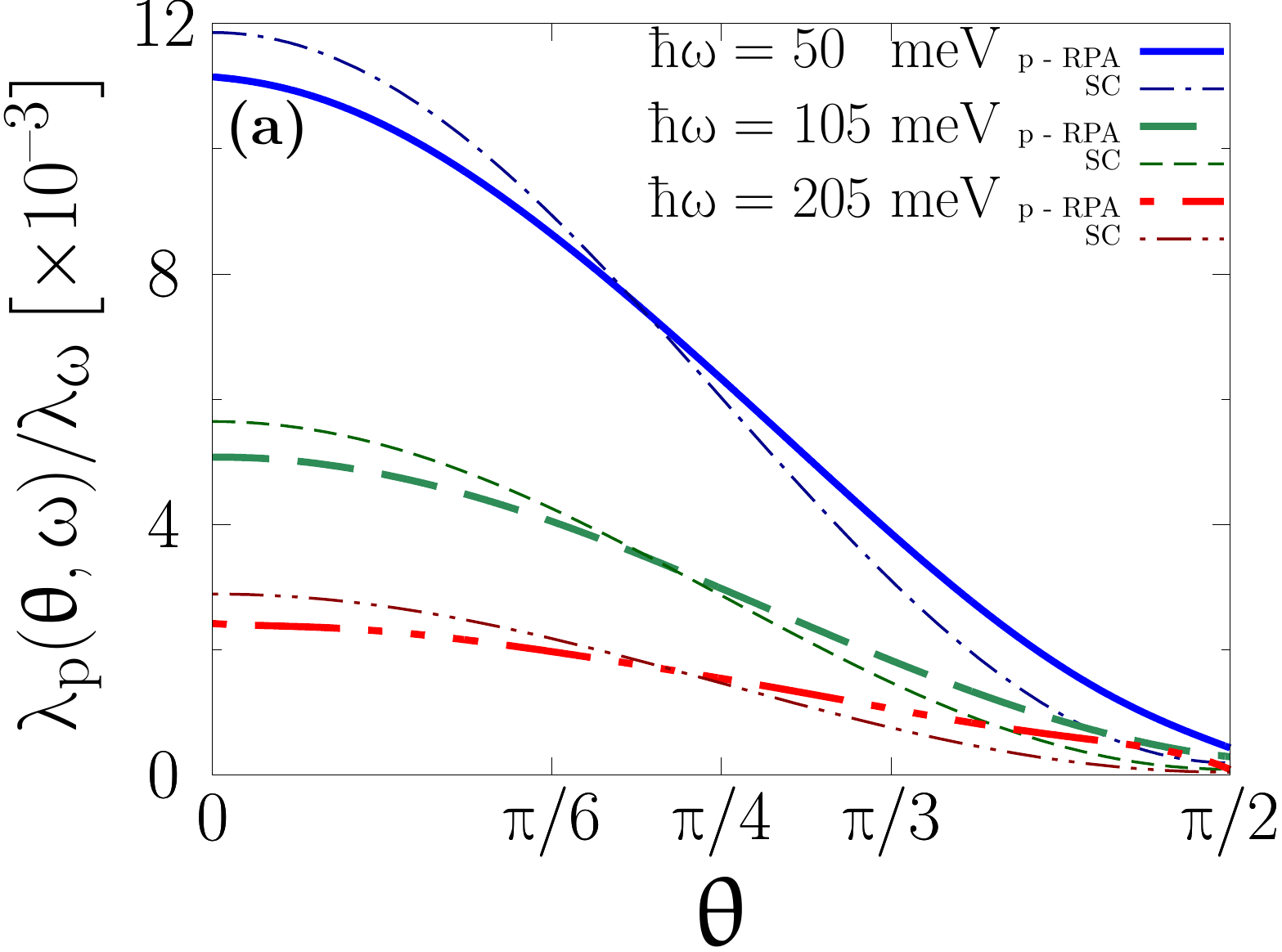}
	   \begin{picture}(0,0)\put(-84,75){\includegraphics[height=2.05cm]{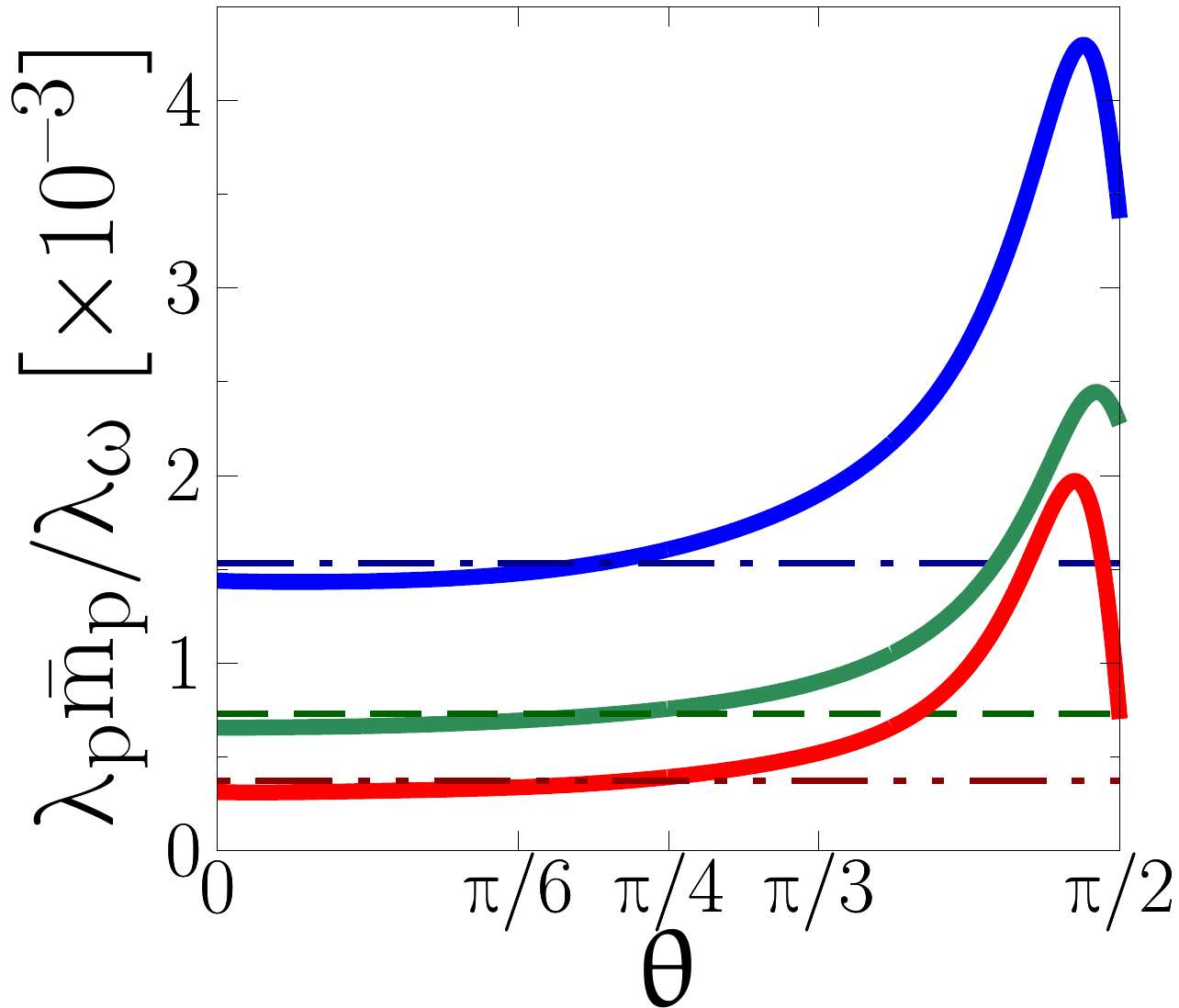}}\end{picture}
	\includegraphics[width=0.492\linewidth]{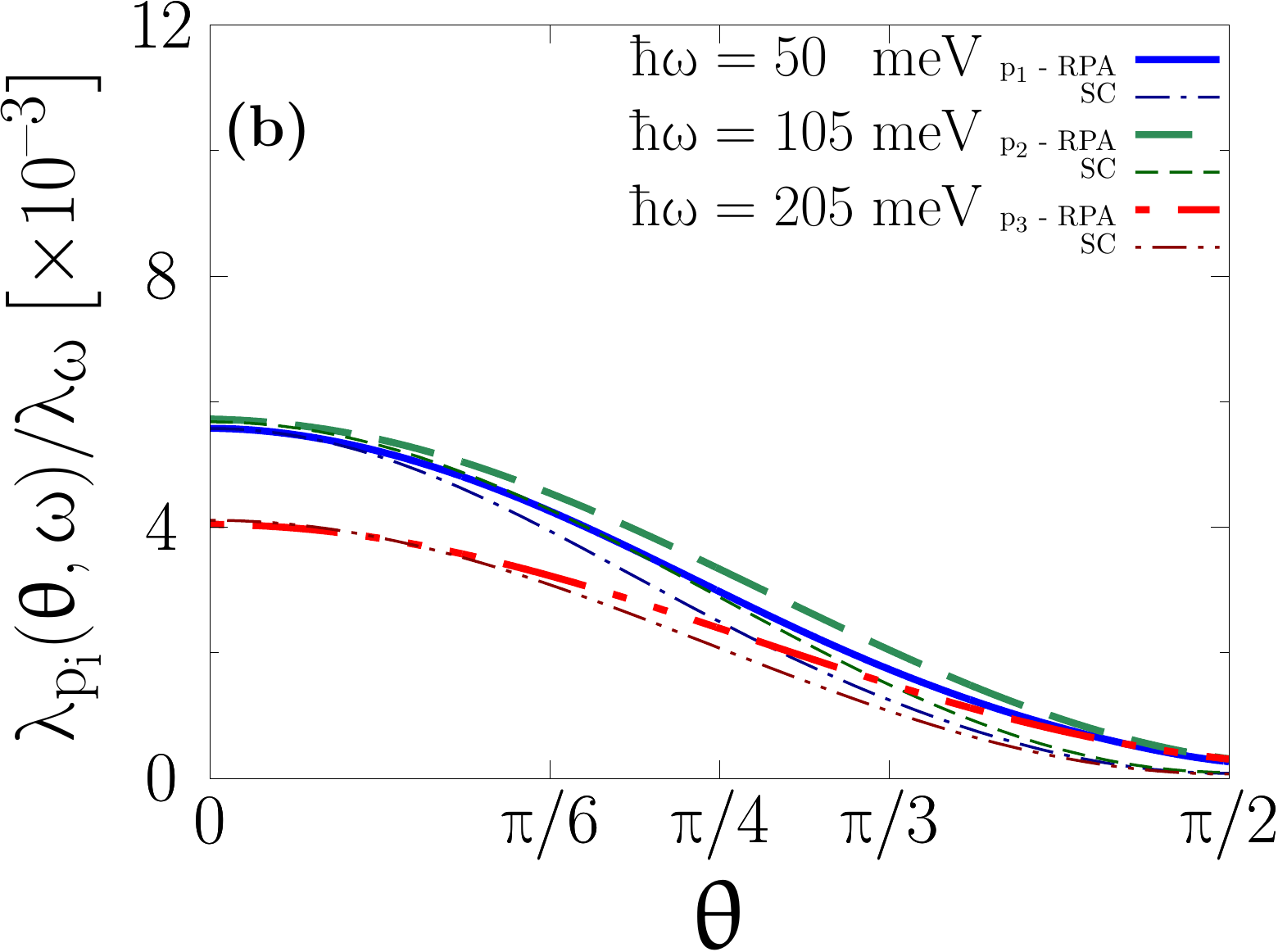}
	   \begin{picture}(0,0)\put(152,66){\includegraphics[height=2.89cm]{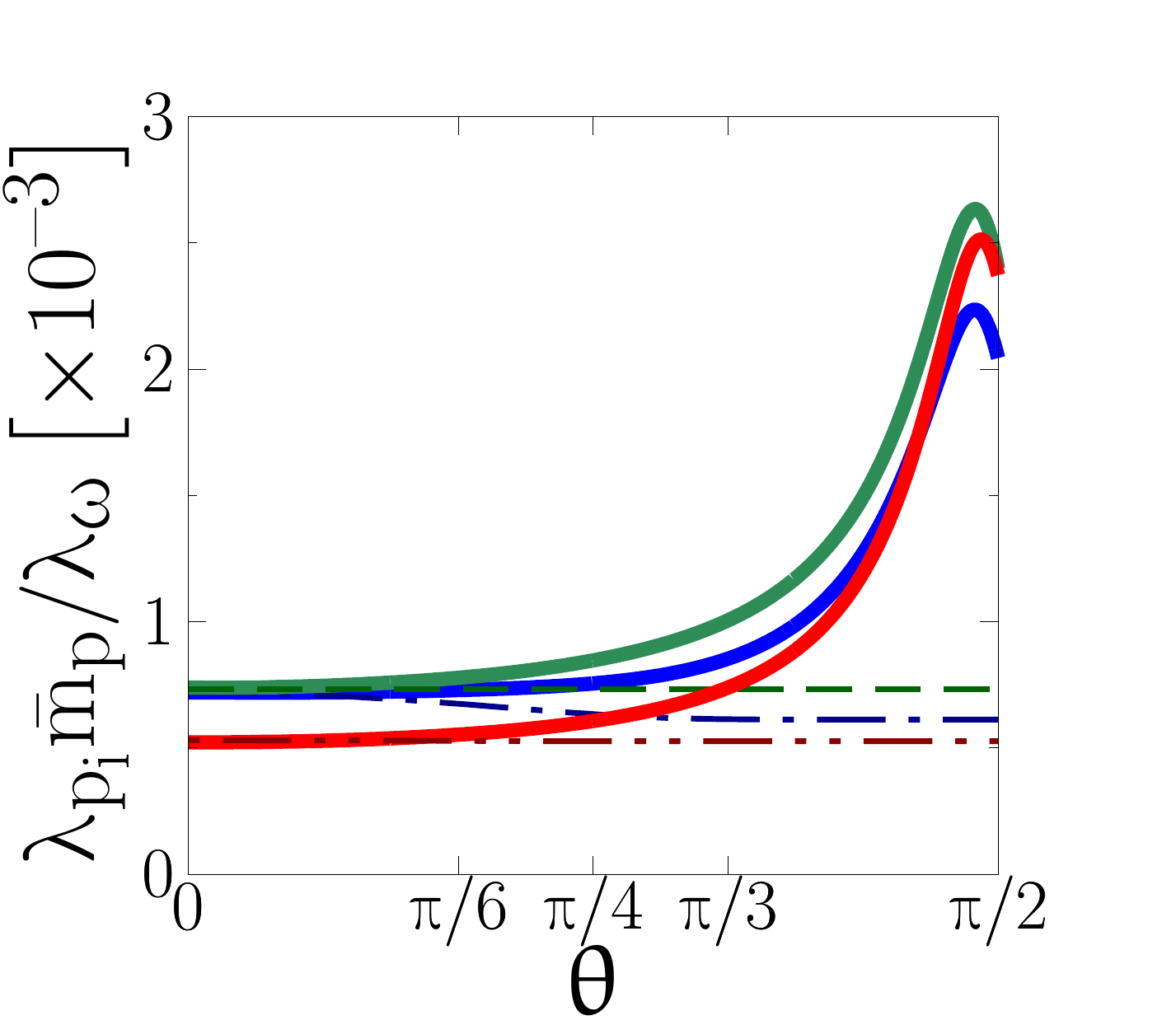}}\end{picture}\\
	\includegraphics[width=0.496\linewidth]{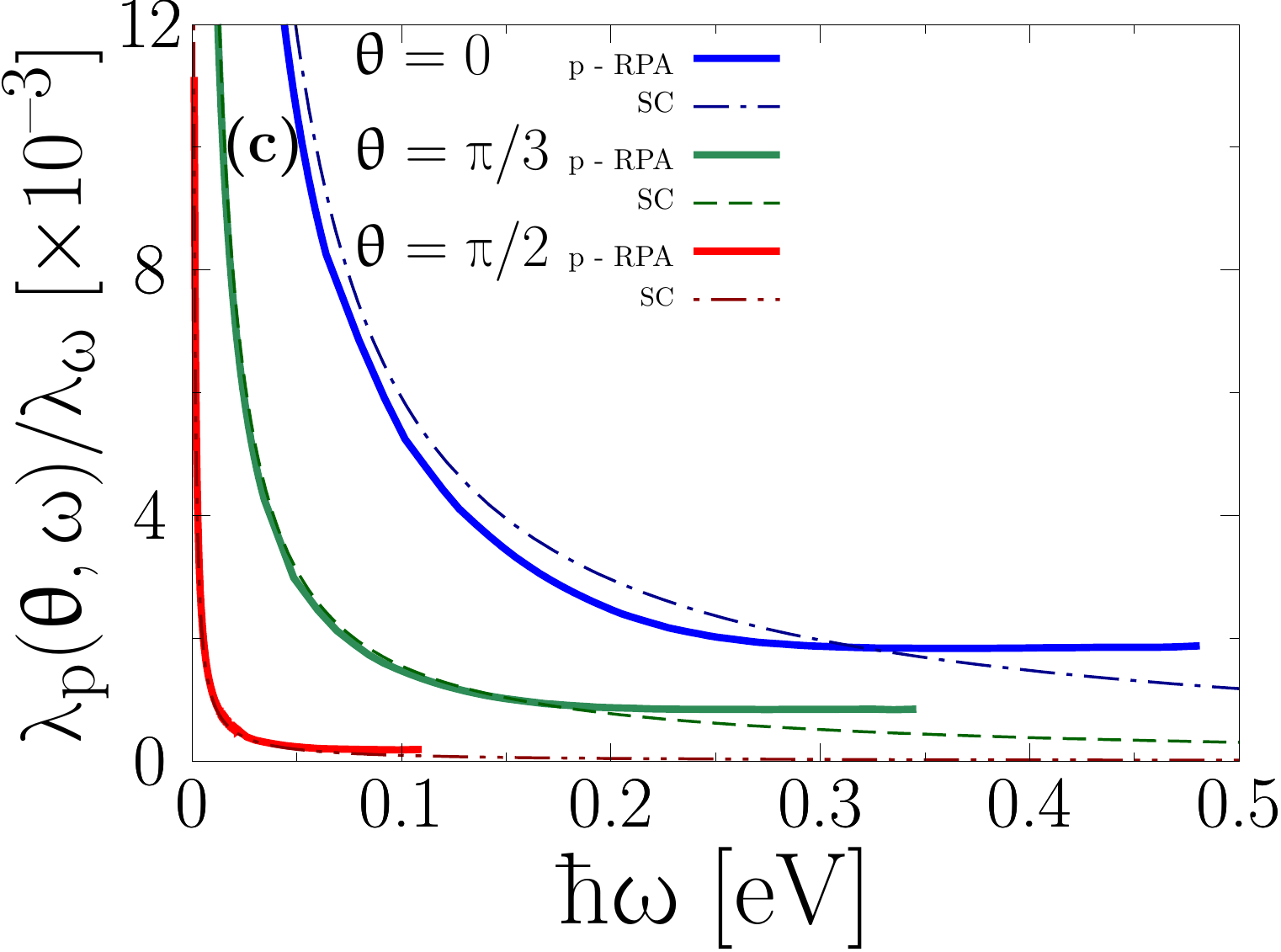}
	   \begin{picture}(0,0)\put(-103,58){\includegraphics[height=2.7cm]{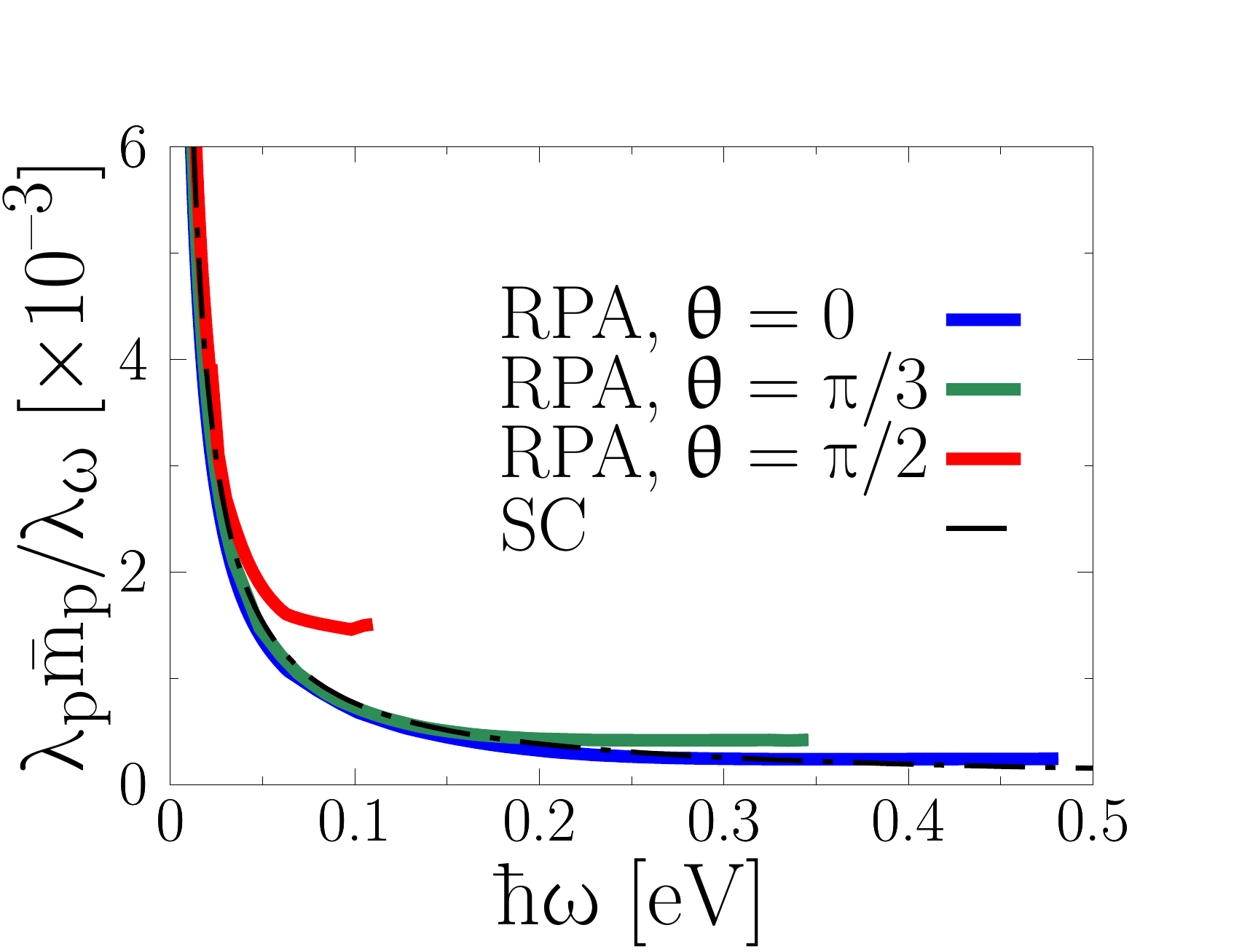}}\end{picture}
	\includegraphics[width=0.49\linewidth]{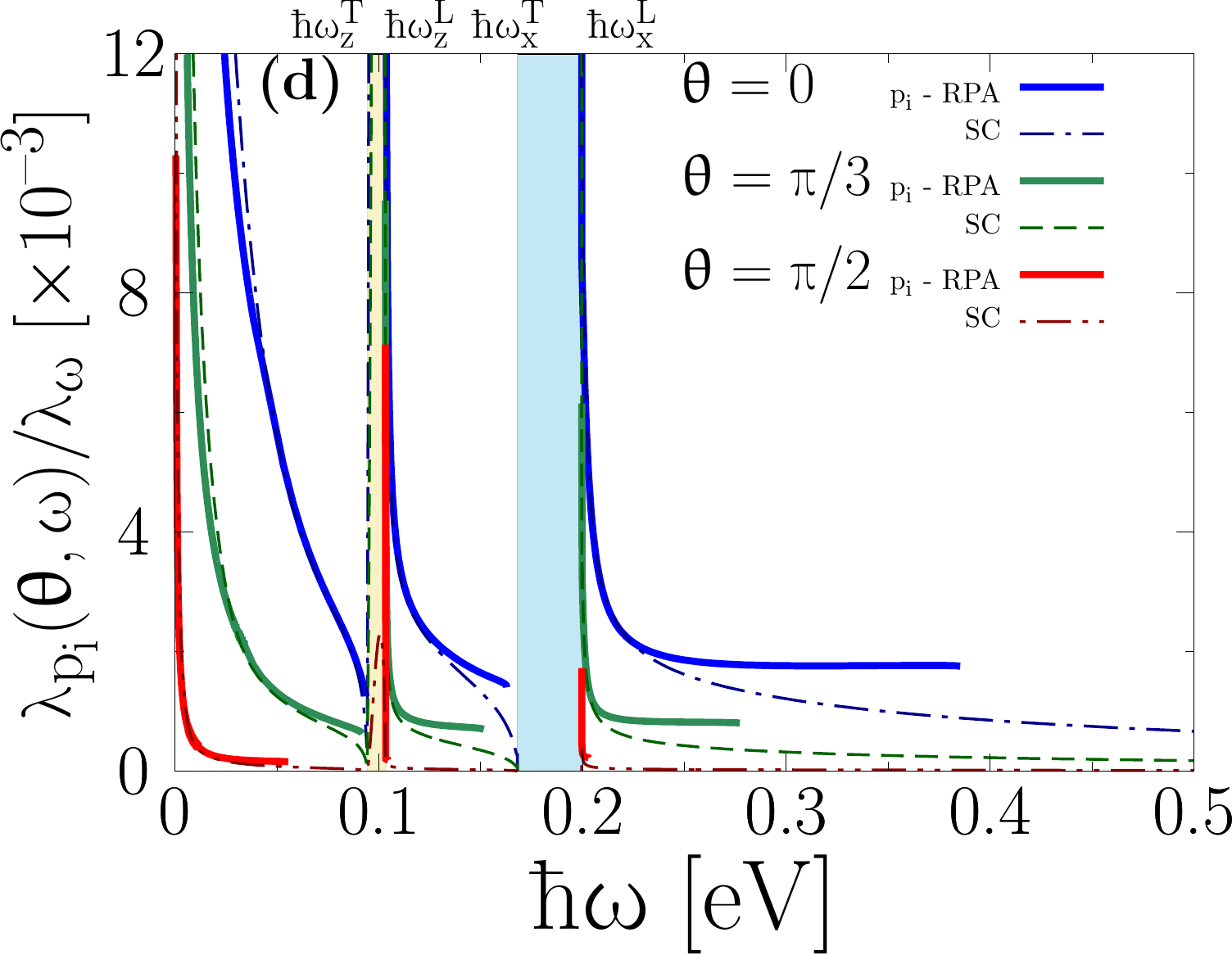}
	   \begin{picture}(0,0)\put(143,68){\includegraphics[height=2.6cm]{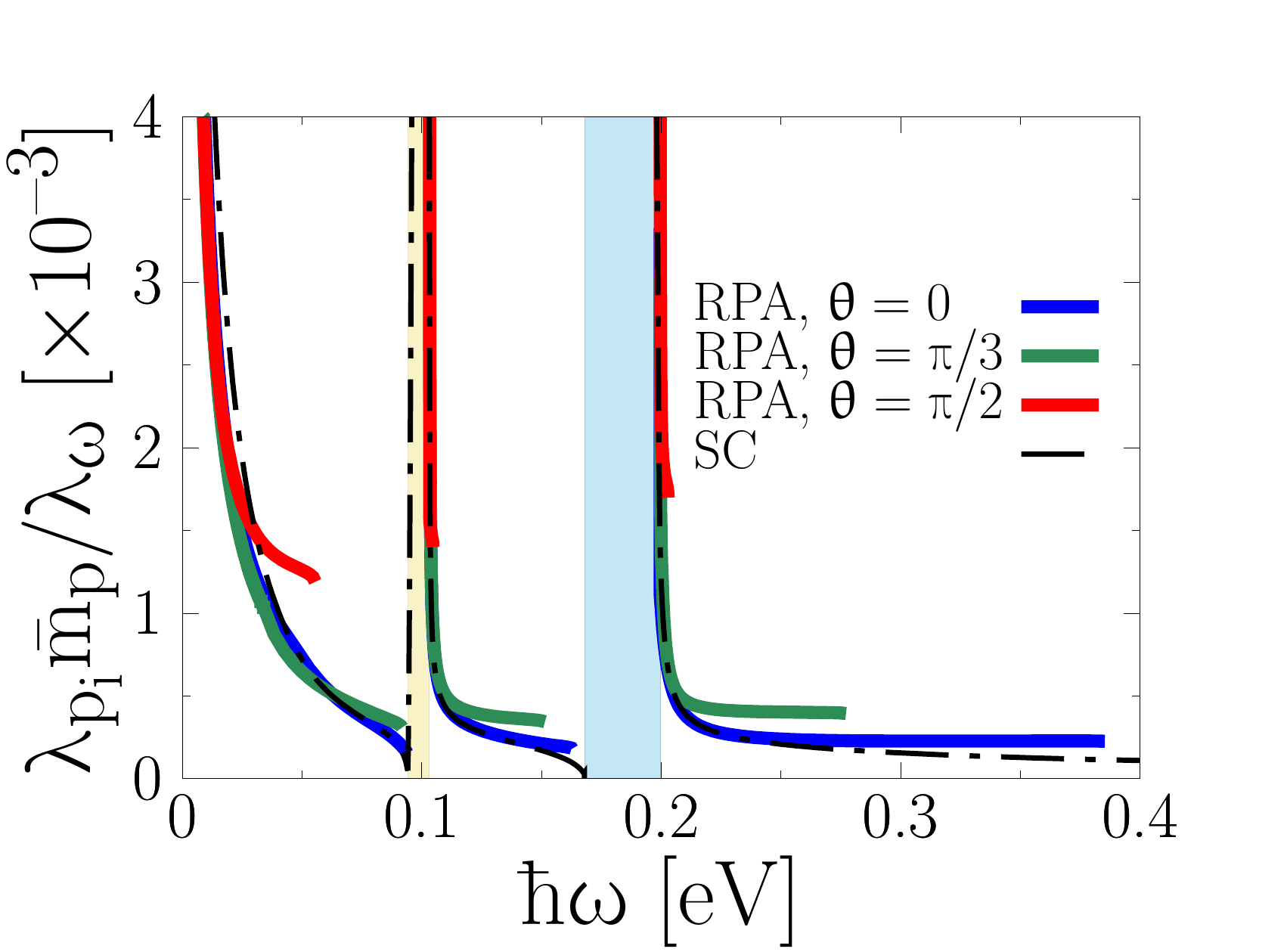}}\end{picture}
	 \caption{(Color online) The scaled lowest plasmon wavelength (relative to the incident light wavelength $\mathrm{\lambda_\omega}=2\pi c/\omega$ emitted from a laser with 3.11 THz frequency) yielded by RPA approach (thick curves) and by the semiclassical (SC) model (thin curves) for the hole doped encapsulated phosphorene at $\mathrm{T=50 \, K}$ with physical conditions described in Fig. \ref{fig6}. Panels (a) and (b) respectively illustrate the simple plasmon scaled wavelength and that of SPPPs as a function of the wavevector azimuthal angle $\mathrm{\theta}$ for different energies $\mathrm{\hbar \omega =50, 105, 205 \; meV}$. Panels (c) and (d) show their dependencies to the frequency $\omega$ along the wavevector directions $\mathrm{\theta=0,\pi/3,\pi/2}$. The insets are the scaled behavior of their corresponding quantities relative to the dimensionless directional mass $\mathrm{\bar{m}_p(\theta)=m_p(\theta)/m_d}$.}
	\label{fig11}
\end{figure*}

Furthermore, in order to have a deeper insight into the nature and behavior of the collective modes and their losses both in the simple and coupled model, we have also calculated their dispersions, propagation lengths and their wavelengths by the semiclassical model based on the Maxwell's classical electrodynamics in the long-wavelength limit $q/k_F \ll 1$ and we have made a comparison between the results of RPA and SC approaches. Making use of the SC dispersion Eqs. (\ref{answer}) and (\ref{answerC}) for the simple surface plasmon and the SPPP modes, in Fig. \ref{fig9} the resulted dispersions of the two approaches are plotted along the wavevector directions (a) $\theta=0$ and (b) $\theta=\pi/2$ for the hole-doped system described before in Fig. (\ref{fig6}) in which the RPA outcomes have been presented initially. The RPA results coincide accurately with that of SC in long-wavelength limit and deviate in higher wavenumbers as expected.

Considering the mentioned dispersion relations in addition to the Eq. (\ref{Lsp}) for the simple plasmon mode's propagation length and it counterpart for the surface palasmon-phonon-polariton modes, in Fig. \ref{fig10} the imaginary part of the wavenumber $\Im m \, q_{p/p_i}(\theta, \omega)$ and the propagation length $L_{p/p_i} (\theta, \omega)$ are depicted as a function of frequency along the wavevector directions $\theta = 0, \pi/3, \pi/2$, and also as a function of wavevector direction $\theta$ for the selected frequencies $\hbar \omega = 50, 105, 205 \,$meV for simple plasmons (panels a, c and e) and also for SPPPs (panels b, d and f). The thick black curves in panels (a-d) are scaled curves relative to the dimensionless directional mass $\bar{m}_p(\theta)=m_p(\theta)/m_d$. The SC approach predicts an anisotropic linear growth of the imaginary part of the simple surface plasmon wavevector by the increasing of the frequency (panel a) and consequently its great loss by a drastic shortening of its propagation length (panel c) that is also intensified by its angular dependency via the term $\bar{m}_p(\theta)$ since it is suppressed heavily by angle increasing up to the zigzag direction $\theta=\pi/2$ (panel e). On the other hand, for the case of SPPP modes, the first gapless mode $p_1$ has almost the same behavior as of the simple plasmon mode in the limit of low energies and the imaginary part of the wavevector for the first two modes increase sharply near the edges of the Reststrahlen bands (panel b), but the third mode has a much lower finite value. These behaviors cause sharp propagation length suppression of the $p_1$ and $p_2$ modes near these frequencies of the surface phonon-polariton modes of the hBN while the $p_3$ mode maintains a moderate propagation length in the vicinity of the $\hbar\omega^L_x$ (panel d). Ultimately, the same aforementioned angular dependency holds also for these modes and by approaching to the zigzag direction the losses dominate and the propagation lengths shrink down (panel f). 

Eventually, according to Eq. (\ref{lambdaP0}) and its counterpart for surface plasmon-phonon-polaritons, the scaled lowest plasmon wavelengths $\lambda_{p/p_i} (\theta, \omega) / \lambda_\omega$ (relative to the incident light wavelength $\lambda_\omega$ emitted from a laser with 3.11 THz frequency) extracted from the RPA and SC approaches (respectively by thick and thin curves) for the aforesaid hole doped encapsulated phosphorene are shown in Fig. \ref{fig11}. The Fig. \ref{fig11}(a) pertains to the simple plasmon wavelength $\lambda_p$ in the simple model as a function of the wavevector angle $\theta$ for the selected energies $\hbar \omega =50, 105, 205\;$meV corresponding to $\hbar\omega_p$. Fig. \ref{fig11}(b) shows the same angular dependency for the SPPPs with energies corresponding to $\hbar \omega_{p_i}$. Figs. \ref{fig11}(c) and \ref{fig11}(d) are respectively pertaining to the frequency dependency of collective mode wavelengths $\lambda_p$ and $\lambda_{p_i}$ along the wavevector directions $\theta=0,\pi/3,\pi/2$. The insets are the scaled behavior of their corresponding quantities relative to the dimensionless directional mass $\bar{m}_p(\theta)=m_p(\theta)/m_d$. The RPA and the SC results are in an agreement on the strong frequency and angular dependency of the scaled wavelength of the collective modes in spite of the fact that the SC overestimates the wavelength shortening by increasing of the wavevector angle $\theta$ and frequency $\omega$. Putting a further step beyond the anisotropic single-band model considered in the SC approach and taking into account the intraband processes and the directional exciton binding hided in the band coupling term $\gamma k_x$ of the Hamiltonian (\ref{Hamiltonian}) would resolve this overestimated blue shift.
Those results uncover some important features of the anisotropic nature of collective modes in encapsulated phosphorene and its relation to practical purposes. As it is appear from the figure, the relative wavelengths decrease drastically by approaching to the zigzag direction $\theta= \pi /2$. Our numerical results predict that the plasmon mode would exhibit higher levels of confinement on the surface along the zigzag direction in the system. This characteristic is the principle for the delivery of light to the nanoscale, including to nano-optical devices, quantum dots and individual molecules \cite{Gramotnev}.
%
\section{Conclusion}
\label{sec4}
%
In summary, we have considered a monolayer phosphorene embedded between two hBN slabs along the heterostructure direction. We have focused on collective modes of a pristine two-dimensional crystalline phosphorene in the presence of different dielectric media. Our studies were based on a low-energy model Hamiltonian together with the random phase approximation.
 
We have calculated the Lindhard function at finite temperature, and then the wavevector azimuthal angle dependence of the plasmon-polariton and plasmon phonon-polariton modes of the hybrid system are obtained. It is essential to emphasize that the material-specific dielectric function for polar substrate are needed to obtain realistic plasmon-phonon dispersions. The lowest plasmon dispersion is displayed as a $\sqrt q$ originating from low-momentum carrier scattering. The anisotropic band structure of monolayer phosphorene along the zigzag and armchair directions make anisotropic features in the collective plasmon excitations, and the plasmon modes in the armchair direction have higher energy compared to the zigzag direction at the same momenta. Furthermore, the collective modes illustrate strong anisotropic and strong coupling with phonon modes of the polar media and moreover, the Landau damping occurs due to the intraband processes when plasmon enters intraband electron-hole continuum.  Our numerical results show that the plasmon mode is highly confined on the surface along the zigzag direction. The phonon dispersions in the Reststrahlen bands are angle-dependent owing to the strong electron-phonon interaction. We have also shown that the energy of the collective modes, which depends also on the Fermi energy, becomes small by increasing the wavevector azimuthal angle and the SPPP modes are somehow phonon-like along the zigzag direction. We have also obtained a semiclassical framework for the long-wavelength limit of the collective behaviors of an anisotropic system. The evaluated results within the semiclassical framework coincide with those of the RPA. Accordingly, the choice of media encapsulated phosphorene can be utilized in order to engineer the plasmon-phonon-polariton dispersions in phosphorene structure.
%
\vspace{-2mm}
\begin{acknowledgements}
We thank M. Polini for very useful discussions. This work is partially supported by Iran Science Elites Federation. R. A acknowledges Scuola Normale Superiore, Pisa for its hospitality during the period in which the final stage of this work is completed.
\end{acknowledgements}
%
%


\begin{thebibliography}{99}
\bibitem{note}
Our numerical results show that the plasmon mode can be fitted with $\sqrt{2\pi n e^2 q/m_{c x}}$ very well along the armchair and with $\sqrt{2\pi n e^2 q/m_{c y}}$ quite good along the zigzag direction at the long-wavelength limit.
\end{thebibliography}
\end{document}